\documentclass[onecollarge]{svjour2}                    % onecolumn
\smartqed 
\usepackage{graphicx}
\usepackage{epsfig,bm}
\usepackage{amsmath,amssymb,psfrag,textcomp}
\usepackage{upgreek} 
\newcommand{\ket}[1]{\left|#1\right\rangle}

\newcommand{\bracket}[1]{\left\langle #1\right\rangle}
\newcommand{\beeq}[1] {\begin{equation}\begin{split}#1\end{split}\end{equation}}
\newcommand{\beeqn}[1] {\begin{equation}\begin{split}#1\nonumber\end{split}\end{equation}}
\begin{document}
\title{A generalised Airy distribution function for the accumulated area swept by $N$ vicious Brownian paths}
\author{Isaac P\'erez Castillo \and Denis Boyer}
\institute{Isaac P\'erez Castillo \and Denis Boyer  \at Departamento de Sistemas Complejos, Instituto de F\'isica, UNAM, P.O. Box 20-364, 01000 M\'exico D.F., M\'exico\\
              \email{isaacpc@fisica.unam.mx}  
          }

\date{Received: date / Accepted: date}
\maketitle
\begin{abstract}
In this work exact expressions for the distribution function of the accumulated area swept  by reunions and meanders of $N$ vicious Brownian particles up to time $T$ are derived. The results are expressed in terms of a generalised Airy distribution function, containing the Vandermonde determinant of the Airy roots. By mapping the problem to an Random Matrix Theory ensemble we are able to perform Monte Carlo simulations finding perfect agreement with the theoretical results.

\keywords{Random walkers \and Vicious walkers \and Random matrices \and Airy distribution function}

\end{abstract}

\section{Introduction}
Since the seminal work of de Gennes on simple models of fibrous structures and its popularisation by Fisher,  the study of vicious walkers (namely, walkers that do not intersect) have attracted attention during the last two decades due  to the wide range of applications in various branches of science  and their connections to random matrix theory \cite{Forrester2011,Schehr2013,Baik2000,Forrester2001,Johansson2003,Nagao2003,Katori2004,Ferrari2005,Tracy2007,Daems2007,Schehr2008,Novak2009,Nadal2009,Rambeau2010,Borodin2010,Bleher2011,Adler2012}. Similarly, the statistical properties of the area swept by a Brownian excursion is a problem that  was originally studied  by mathematicians. The Laplace transform of the distribution of the area (known as the Airy distribution function) was first computed by Darling and Louchard \cite{Darling1983,Louchard1984}. The derivation of its moments together with  the actual distribution were obtained by Tak\'acs \cite{Takacs1991,Takacs1993,Takacs1995}. The Airy distribution function has appeared in a number of problem from different areas from  graph theory and computer science as well as in  physical systems modelling one dimensional fluctuating interfaces \cite{Majumdar2004,Majumdar2005}, applications in laser cooling \cite{Barkai2014,Kessler2014} and fluctuations of sizes of ring polymers \cite{Medalion2015}.

In this work, we focus on obtaining exact expressions for the accumulated area swept by $N$ vicious Brownian motions. This paper is organised as follows: in section \ref{sec:MD}, we define the process of vicious walkers and the stochastic quantities we are interested in. In particular, we show in section \ref{sec:PIA} that the Laplace transform of PDF of the accumulated area  can be expressed as a ratio of two N-particle propagators, which can be constructed as Slater determinants of single-particle states. This is a straightforward consequence of using Quantum Mechanics formalism or, alternatively,  of the Karlin-McGregor formula. The resulting expression is fairly general. In order to extract some properties we consider two type of processes: reunions and meanders. Moreover we consider two type of boundary conditions at $x=0$: absorbing boundary conditions (section \ref{sec:ABCs}) and reflecting boundary conditions (section \ref{sec:RBCs}). The inverse Laplace transform is resolved  in section \ref{sec:ILT} and a formula for its negative moments is obtained in  section \ref{sec:M}. Finally, in section \ref{sec:MCS}, using a mapping  between processes of vicious walkers to ensembles of Random Matrix Theory we are able to perform Monte Carlo simulations and compare with the exact formulas obtained in the previous sections.

\section{Model definitions}
\label{sec:MD}
Consider the set of trajectories defined by $N$ one-dimensional Brownian particles $\bm{x}(\tau)=(x_1(\tau),\ldots,x_{N}(\tau))$ with $\tau\in[0,T]$ such that $0<x_1(\tau)<\cdots<x_N(\tau)$ and let us denote as $\mathcal{M}_{N}[\bm{x}(\tau)]$ its probability measure. Next, let us define as $A_i$ the area swept by particle $i$:
\beeqn{
A_i=\int_0^T x_i(\tau)d\tau\,.
}
The set of areas $\bm{A}=(A_1,\ldots,A_N)$ is a random $N$-tuple whose joint PDF (jPDF) is given by \textit{viz.}
\beeq{
\mathcal{P}_{N}(\bm{A},T)=\int_{\bm{x}(0)=\bm{x}_{i}}^{\bm{x}(T)=\bm{x}_f} \mathcal{D}\bm{x}(\tau)\mathcal{M}_N[\bm{x}(\tau)]\prod_{i=1}^N\delta\left(\int_0^Tx_i(\tau)d\tau-A_i\right)\,.
\label{eq:jpdf}
}
Notice that the jPDF \eqref{eq:jpdf} contains the PDF of the area swept for each curve $i=1,\ldots,N$ 
\beeq{
\mathcal{P}^{(i)}_{N}(A_i,T)=\int_0^{\infty}\cdots\int_0^{\infty} dA_1\cdots dA_{i-1}dA_{i+1}\cdots dA_{N} \mathcal{P}_{N}(\bm{A},T)\,,
\label{eq:inner}
}
and, in particular, the PDFs of the area for the bottom and top curves denoted as $\mathcal{T}_{N}(A,T)\equiv\mathcal{P}^{(N)}_{N}(A,T)$ and $\mathcal{B}_{N}(A,T)\equiv \mathcal{P}^{(1)}_{N}(A,T)$, respectively. It also contains the PDF of the total  area, defined as
\beeq{
\mathcal{S}_N(A,T)=\int_0^{\infty}\cdots\int_0^{\infty}dA_1\cdots dA_{N} \mathcal{P}_{N}(\bm{A},T)\delta\left(A-\sum_{i=1}^NA_i\right)\,.
\label{eq:tsa}
}
In this work we present exact formulas for the PDF  $\mathcal{S}_N(A,T)$ for the accumulated area and compare the expressions with Monte Carlo simulations. A thorough analysis based on Monte Carlo simulations on  the marginals $\mathcal{P}^{(i)}_{N}(A,T)$ for $i=1,\ldots,N$ will be discussed somewhere else.

\section{Path Integral Approach}
\label{sec:PIA}
Combining \eqref{eq:jpdf} and \eqref{eq:tsa} yields
\beeqn{
\mathcal{S}_N(A,T)=\int_{\bm{x}(0)=\bm{x}_i}^{\bm{x}(T)=\bm{x}_f} \mathcal{D}\bm{x}(\tau)\mathcal{M}_N[\bm{x}(\tau)]\delta\left(A-\sum_{i=1}^{N}\int_0^T  x_{i}(\tau)d\tau\right)\,,
}
with
\beeqn{
\mathcal{M}[\bm{x}(\tau)]=\frac{1}{Z}\exp\left[-\frac{1}{2}\sum_{i=1}^N\int_0^{T}d\tau\left(\frac{dx_i(\tau)}{d\tau}\right)^2\right]C_N[\bm{x}(\tau)]\,,
}
where $Z$ is the normalisation factor and $C_{N}[\bm{x}(\tau)]$ is an indicator function imposing the constraint of vicious particles  $0<x_1(\tau)<\cdots<x_N(\tau)$ on the positive part of the real line. Further, the Laplace transform of $\mathcal{S}_{N}(A,T)$,
\beeqn{
\widehat{\mathcal{S}}_{N}(\lambda,T)=\bracket{e^{-\lambda A}}_{\mathcal{S}_N}\equiv\int_0^{\infty} dA~\mathcal{S}_{N}(A,T)e^{-\lambda A}
}
has the following representation using path integral approach:
\beeq{
\widehat{\mathcal{S}}_{N}(\lambda,T)&=\int^{\bm{x}(T)=\bm{x}_f}_{\bm{x}(0)=\bm{x}_i}\mathcal{D}\bm{x}(\tau)\mathcal{M}_N[\bm{x}(\tau)]\exp\left[-\lambda\sum_{i=1}^N\int_0^{T} x_i(\tau)d\tau\right]\,.
\label{eq:lt}
}
In Quantum Mechanics formalism, the expression \eqref{eq:lt} can be expressed as the ratio of two propagators
\beeq{
\widehat{\mathcal{S}}_{N}(\lambda,T)&=\frac{G^{(1)}_N(\bm{x}_f,T|\bm{x}_i,0)}{G_N^{(0)}(\bm{x}_f,T|\bm{x}_i,0)}\,,
\label{eq:lt2}
}
with $G^{(a)}_N(\bm{x},T|\bm{x}_0,0)=\bracket{\bm{x}|e^{-H^{(a)}_N T}|\bm{x}_0}$ and Hamiltonian operators
\beeq{
H^{(a)}_N=\sum_{j=1}^N h_j^{(a)}=-\frac{1}{2}\sum_{j=1}^N\partial_{x_j}^2+\sum_{j=1}^NV^{(a)}(x_j)\,,\quad\quad a=0,1
}
and $V^{(0)}(x)$ a confining potential in $\mathbb{R}^{+}$, while $V^{(1)}(x)=\lambda x$ for $x>0$ and infinity for $x\leq 0$. Besides, Brownian particles are not allowed to cross which in QMf corresponds to having fermions. Considering the latter and that the potentials are one-particle type operators, the eigenfunctions of $H^{(a)}_N$ are given by the Slater determinant of the eigenfunctions corresponding to the one-particle problem $h^{(a)}$, that is
\beeq{
G_N^{(a)}(\bm{y},T|\bm{x},0)&=\sum_{\bm{n}}\Phi^{(a)}_{\bm{n}}(\bm{y})\overline{\Phi}^{(a)}_{\bm{n}}(\bm{x})e^{-E^{(a)}_{\bm{n}} T}\,,
\label{eq:prop}
}
with
\beeqn{
\Phi^{(a)}_{\bm{n}}(\bm{x})=\frac{1}{\sqrt{N!}}\det_{1\leq i,j\leq N}\phi^{(a)}_{n_i}(x_j)\,,\quad\quad E^{(a)}_{\bm{n}} =\sum_{i=1}^N e^{(a)}_{n_i}\,,\quad\quad h^{(a)}\ket{\phi^{(a)}_n}=e^{(a)}_{n}\ket{\phi^{(a)}_n}\,.
}
Once the corresponding one-particle problems of the numerator and denominator have been solved, we have solved the problem via the formula \eqref{eq:lt2}. However, the corresponding expression is still too general to extract general mathematical properties. Thus, in light of the previous work we will particularise \eqref{eq:lt2} to the cases of  reunions and  meanders.\\
 For reunions we mean the set of processes by which the $N$ vicious Brownian paths start and finish at the same point, that is $\bm{x}_i=\bm{x}_f=\bm{0}$. This is achieved by taking $\bm{x}_i=\bm{x}_f=\bm{\epsilon}$  with $\bm{\epsilon}=(\epsilon_1,\cdots,\epsilon_N)$, and taking the limit $\bm{\epsilon}$ to $\bm{0}$, \textit{viz.}
\beeq{
\widehat{\mathcal{S}}^{(r)}_{N}(\lambda,T)&=\lim_{\bm{\epsilon}\to\bm{0}}\frac{G^{(1)}_N(\bm{\epsilon},T|\bm{\epsilon},0)}{G_N^{(0)}(\bm{\epsilon},T|\bm{\epsilon},0)}\,.
\label{eq:reunion}
}
For meanders we mean the set of processes where the paths start at $\bm{x}_i=\bm{0}$ and are allowed to finish at any final position $\bm{x}_f$ at time $T$. As such we need to integrate over all possible final positions and, therefore, the Laplace transform of the corresponding PDF reads
\beeq{
\widehat{\mathcal{S}}^{(m)}_{N}(\lambda,T)=\lim_{\bm{\epsilon}\to\bm{0}}\frac{\int_{\bm{W}_N} d^N\bm{x}_f~G^{(1)}_N(\bm{x}_f,T|\bm{\epsilon},0)}{\int_{\bm{W}_N} d^N\bm{x}_f~G_N^{(0)}(\bm{x}_f,T|\bm{\epsilon},0)}\,,
\label{eq:meander}
}
where  $\bm{W}_N$ stands for the Weyl chamber $\bm{W}_N=\left\{\bm{x}\in\mathbb{R}^{N}|0\leq x_1\leq \cdots \leq x_N\right\}$ and $d^{N}\bm{x}=dx_1\cdots dx_N$.

\section{Absorbing Boundary conditions}
\label{sec:ABCs}
We note in the two cases above that we will need the one-particle problem for $a=1$ (the one-particle problem for $a=0$ is treated in the appendices). For absorbing boundary conditions (i.e. the corresponding eigenfunctions of the one-particle problem obey $\phi(x=0)=0$) it has the well-known solution
\beeqn{
\phi^{(1)}_{n}(x)=\frac{\sqrt{(2\lambda)^{1/3}}}{|\text{Ai}'(-\alpha_{n})|}\text{Ai}[(2\lambda)^{1/3}x-\alpha_{n}]\,,\quad e^{(1)}_{n}=\alpha_{n}\lambda^{2/3}2^{-1/3}\,,\quad\quad  n=1,2,\ldots
}
where $\text{Ai}(x)$ is the Airy function with zeros $z=-\alpha_n$ with $n\in\{1,2,\ldots\}$. Thus the eigenfunction for the $N$-particle problem is:
\beeq{
\Phi^{(1)}_{\bm{n}}(\bm{x})&=\frac{1}{\sqrt{N!}}\det_{1\leq i,j\leq N}\phi^{(1)}_{n_i}(x_j)=\frac{(2\lambda)^{N/6}}{\sqrt{N!}\prod_{i=1}^N|\text{Ai}'(-\alpha_{n_i})|}\det_{1\leq i,j\leq N}\left[\varphi_{n_i}(z_j)\right]\,,
\label{eq:sd}
}
where we have defined $\varphi_{n}(z)=\text{Ai}(z-\alpha_{n})$ with $z=(2\lambda)^{1/3}x$.
\subsection{Reunions}
To obtain an exact expression for $\widehat{\mathcal{S}}^{(r)}_{N}(\lambda,T)$ given by \eqref{eq:reunion} we need to perform the Taylor expansion of the Slater determinant \eqref{eq:sd} containing the Airy eigenfunctions. To do so we make use of the following formula \cite{Laurenzi2011}:
\beeqn{
\text{Ai}(t)\text{Bi}(z+t)-\text{Bi}(t)\text{Ai}(z+t)=\frac{1}{\pi}\sum_{\ell=0}^\infty\frac{z^\ell}{\ell!}Q_\ell(t)\,,
}
with $Q_n(z)$ the Abramochkin polynomials \cite{Laurenzi2011}. In our case, we take that $t=-\alpha_n$ which yields
\beeqn{
\text{Ai}(z-\alpha_n)=\text{Ai}'(-\alpha_n)\sum_{\ell=0}^\infty\frac{z^\ell}{\ell!}Q_\ell(-\alpha_n)\,,
}
where we have used the identity $\pi\text{Bi}(-\alpha_n)\text{Ai}'(-\alpha_n)=-1$. Combining this together with the standard formula of the Taylor series of a Slater determinant (see appendix \ref{ap:sd}), we arrive at
\beeq{
\det_{1\leq i,j\leq N}\varphi_{n_i}(z_j)&=\left[\prod_{j=1}^N\text{Ai}'(-\alpha_{n_j})\right]\sum_{\bm{\ell}\in U_N}\frac{(2\lambda)^{\frac{1}{3}\sum_{j=1}^N\ell_j}}{\ell_1!\cdots \ell_N!}\det_{1\leq i,j\leq N}[x^{\ell_j}_i]\det_{1\leq i,j\leq N}[Q_{\ell_i}(-\alpha_{n_j})]\,,
\label{eq:a}
}
where $\bm{\ell}\in U_{N}$ stands for the sum over all ordered $N$-tuple indices $\bm{\ell}=(\ell_1,\ldots,\ell_N)$, that is with $\ell_1<\ell_2<\cdots<\ell_N$.\\
Everything boils down to  be able to find an expression for the determinant $\det_{1\leq i,j\leq N}[Q_{\ell_i}(-\alpha_{n_j})]$ of the Abramochkin polynomials evaluated at the Airy roots. It turns out that it is possible  to obtain the lowest contribution in $\bm{\epsilon}$ if we recall that the odd-numbered polynomials are such that $Q_{2n+1}(z)= z^{n}+\text{lower order terms}$, while even-numbered polynomials can be expressed in terms of the previous odd-numbered ones. This automatically implies that the first non-zero lowest-order contribution in $\bm{\epsilon}$ in the sum in eq. \eqref{eq:a} is given by taking $\ell_j=2(j-1)+1$ for $j=1,\ldots,N$. This eventually yields:
\beeq{
\det_{1\leq i,j\leq N}\varphi_{n_i}(z_j)&\sim (2\lambda)^{\frac{N^2}{3}}\left[\prod_{j=1}^N\text{Ai}'(-\alpha_{n_j})\right]\left[\prod_{i=1}^N x_i\right]\Delta_N(x_1^2,\ldots x_N^2)\Delta_N(\alpha_{n_1},\dots \alpha_{n_N})\,,
\label{eq:a2}
}
where we have defined $\gamma_N=(-1)^{\frac{N(N-1)}{2}}/[1!\cdots (2N+1)!]$, and $\Delta_N(a_1,\ldots,a_N)$ is the Vandermonde determinant:
\beeqn{
\Delta_N(a_1,\ldots,a_N)\equiv\begin{vmatrix}
1&\cdots&1\\
a_1&\cdots&a_N\\
\vdots&\ddots&\vdots\\
a_{1}^{N-1}&\cdots& a_{N}^{N-1}\\
\end{vmatrix}=\prod_{1\leq i<j\leq N}(a_i-a_j)\,.
}
To obtain the expression \eqref{eq:a2} we have used that
\beeqn{
\det_{1\leq i,j\leq N}[x^{2(j-1)+1}_i]&=\left[\prod_{i=1}^N x_i\right]\Delta_N(x_1^2,\ldots x_N^2)\,,\quad\det_{1\leq i,j\leq N}[Q_{\ell_i}(-\alpha_{n_j})]=(-1)^{\frac{N(N-1)}{2}}\Delta_N(\alpha_{n_1},\dots \alpha_{n_N})\,.
}
The final expression of the Taylor expansion for the $N$-particle eigenfunction is:
\beeq{
\Phi^{(1)}_{\bm{n}}(\bm{\epsilon})&\sim \beta_N(2\lambda)^{\frac{N(2N+1)}{6}}\left[\prod_{j=1}^N\frac{\text{Ai}'(-\alpha_{n_j})}{|\text{Ai}'(-\alpha_{n_j})|}\right]\left[\prod_{i=1}^N \epsilon_i\right]\Delta_N(\epsilon_1^2,\ldots \epsilon_N^2)\Delta_N(\alpha_{n_1},\dots \alpha_{n_N})\,,
\label{eq:dte}
}
with $\beta_N=\gamma_N/\sqrt{N!}$. A similar derivation can be done for the denominator (see appendix \ref{ap:nf}). Using the results \eqref{eq:dte},  \eqref{eq:propapp}, and \eqref{eq:39} in the corresponding expression \eqref{eq:prop} of the propagator, and after rearranging terms, we eventually obtain the Laplace transform of the PDF $\mathcal{S}^{(r)}_N(A,T)$, \textit{viz.}
\beeq{
\widehat{\mathcal{S}}^{(r)}_N(\lambda,T)&=\frac{\left(\lambda T^{3/2}\right)^{E^{(r)}_N}}{A^{(r)}_{N}}\sum_{n_1,\ldots,n_N} \Delta^2_N(\alpha_{n_1},\dots \alpha_{n_N})e^{-(\lambda T^{3/2})^{2/3}2^{-1/3}\sum_{i=1}^N\alpha_{n_i}}\,,\\
A^{(r)}_N&=\frac{2^{\frac{1}{2} E_N^{(r)}}}{\pi^N}\prod_{j=0}^{N-1}\Gamma(2+j)\Gamma(3/2+j)\,,\quad\quad E^{(r)}_N=\frac{1}{3}N(2N+1)\,.
\label{eq:ltpdf1}
}
\subsection{Meanders}
For the case \eqref{eq:meander} we report the final result (the derivation can be found in appendix \ref{app:meander})
\beeq{
\widehat{\mathcal{S}}^{(m)}_N(\lambda,T)&=\frac{\left(\lambda T^{3/2}\right)^{E^{(m)}_N}}{A^{(m)}_{N}}\sum_{n_1,\ldots,n_N} B^{(A)}_N(\alpha_{n_1},\ldots,\alpha_{n_N})\Delta_N(\alpha_{n_1},\dots \alpha_{n_N})e^{-(\lambda T^{3/2})^{2/3}2^{-1/3}\sum_{i=1}^N\alpha_{n_i}}\,,
\label{eq:ltpdf2}
}
where 
\beeqn{
B^{(A)}_N(\alpha_{n_1},\ldots,\alpha_{n_N})&=\frac{1}{\prod_{i=1}^N\text{Ai}'(-\alpha_{n_i})}\int_{\bm{W}_N} d^N\bm{x}\det_{1\leq i,j\leq N}\left[\text{Ai}(x_j-\alpha_{n_i})\right]\,,\quad E^{(m)}_N=\frac{N^2}{3}\,.
}
In this case we were unable to find a simple expression for the normalisation constant $A_N^{(m)}$.

\section{Reflecting boundary conditions}
\label{sec:RBCs}
So far we have preoccupied ourselves by considering absorbing boundary conditions at $x=0$. We have also considered how the preceding derivations change when we take reflecting boundary conditions instead. Apart from the mathematical curiosity,  the resulting  process is interesting as it is related to the work \cite{Katori2004} for processes $\bm{X}^{(-1/2,0)}(t)$ which show a transition from type D to type D' matrix ensembles (see \cite{Katori2004} for details).\\
The first thing is to notice how the one-particle wave-function changes in this case. If we denote as $z=-\beta_n$ with $n\in{1,2,\ldots}$ the zeros of the derivative of the Airy function $\text{Ai}'(x)$, then
\beeqn{
\phi_{n}^{(1)}(x)=\frac{\sqrt{(2\lambda)^{1/3}}}{\sqrt{\beta_n}|\text{Ai}(-\beta_n)|}\text{Ai}((2\lambda)^{1/3}x-\beta_n)\,,\quad\quad e^{(1)}_n=\beta_n \lambda^{2/3}2^{-1/3}
}
Thus, in this case the $N$-particle wavefunction takes the following form:
\beeqn{
\Psi^{(1)}_{\bm{n}}(\bm{x})=\frac{(2\lambda)^{N/6}}{\sqrt{N!}\prod_{i=1}^N\sqrt{\beta_{n_i}}|\text{Ai}(-\beta_{n_i})|}\det_{1\leq i,j\leq N}[\psi_{n_i}(z_j)]\,,\quad\quad \psi_{n}(z)=\text{Ai}(z-\beta_n)\,,\quad z=(2\lambda)^{1/3}x\,.
}
\subsection{Reunions}
We are left with deriving a Taylor expansion of the Slater determinant, as before. In this case, the other Abramochkin polynomials \cite{Laurenzi2011} become useful:
\beeqn{
\text{Bi}'(z)\text{Ai}(z+t)-\text{Ai}'(z)\text{Bi}(z+t)=\frac{1}{\pi}\sum_{n=0}^{\infty}\frac{t^{n}}{n!}P_n(z)\,.
}
From this expression we choose $t=-\beta_n$ to write
\beeqn{
\text{Ai}(z-\beta_n)=\text{Ai}(-\beta_n)\sum_{\ell=0}^{\infty}\frac{t^{\ell}}{\ell!}P_\ell(-\beta_\ell)\,,
}
where we have used the property $\pi\text{Bi}'(-\beta_n)\text{Ai}(-\beta_n)=1$. The expansion of the Slater determinant yields
\beeqn{
\det_{1\leq i,j\leq N}\psi_{n_i}(z_j)&=\left[\prod_{j=1}^N\text{Ai}(-\beta_{n_j})\right]\sum_{\bm{\ell}\in U_N}\frac{(2\lambda)^{\frac{1}{3}\sum_{j=1}^N\ell_j}}{\ell_1!\cdots \ell_N!}\det_{1\leq i,j\leq N}[x^{\ell_j}_i]\det_{1\leq i,j\leq N}[P_{\ell_i}(-\beta_{n_j})]\,,
}
Again here, the lowest contribution in $\bm{\epsilon}$ is given by the even-labelled polynomials $P_{2n}(z)=z^n+\cdots$ for $n=0,1,2\ldots$ and with $P_0(z)=1$. Thus we must take $\ell_j=2(j-1)$ for $j=1,\ldots,N$ yielding the following result for the Slater determinant
\beeqn{
\Psi^{(1)}_{\bm{n}}(\bm{\epsilon})=\delta_N(2\lambda)^{\frac{N(2N-1)}{6}}\left[\prod_{j=1}^N\frac{\text{Ai}(-\beta_{n_j})}{\sqrt{\beta_{n_j}}|\text{Ai}(-\beta_{n_j})|}\right]\Delta_{N}(x_1^2,\ldots,x_N^2)\Delta_N(\beta_{n_1},\ldots,\beta_{n_N})
}
with $\delta_N=(-1)^{\frac{N(N-1)}{2}}/[0!2!\cdots 2(N-1)!\sqrt{N!}]$. A similar analysis can be done to the propagator in the denominator of eq. \eqref{eq:lt2} (see appendix \ref{ap:nf}) eventually obtaining
\beeqn{
\widehat{\mathcal{S}}^{(r,R)}_N(\lambda,T)&=\frac{\left(\lambda T^{3/2}\right)^{E^{(r)}_N}}{A^{(r)}_{N}}\sum_{n_1,\ldots,n_N} \frac{\Delta^2_N(\beta_{n_1},\dots, \beta_{n_N})}{\beta_{n_1}\cdots\beta_{n_N}}e^{-(\lambda T^{3/2})^{2/3}2^{-1/3}\sum_{i=1}^N\beta_{n_i}}\\
A^{(r)}_N&=\frac{2^{\frac{1}{2}E^{(r)}_N}}{\pi^N}\prod_{j=0}^{N-1}\Gamma\left(\frac{1}{2}+j\right)\Gamma\left(j+2\right)\,,\quad\quad E^{(r)}_N=\frac{1}{3}N(2N-1)\,,\quad\quad 
}

\subsection{Meanders}
Similarly, in the case  of meanders (see appendix \ref{app:meander}), we obtain
\beeq{
\widehat{\mathcal{S}}^{(m)}_N(\lambda,T)&=\frac{\left(\lambda T^{3/2}\right)^{E^{(m,R)}_N}}{A^{(m)}_{N}}\sum_{n_1,\ldots,n_N} B^{(R)}_N(\beta_{n_1},\ldots,\beta_{n_N})\frac{\Delta_N(\beta_{n_1},\dots \beta_{n_N})}{\beta_{n_1}\cdots\beta_{n_N}}e^{-(\lambda T^{3/2})^{2/3}2^{-1/3}\sum_{i=1}^N\beta_{n_i}}\,,
\label{eq:ltpdf2}
}
where 
\beeqn{
B^{(R)}_N(\beta_{n_1},\ldots,\beta_{n_N})&=\frac{1}{\prod_{i=1}^N\text{Ai}(-\beta_{n_i})}\int_{\bm{W}_N} d^N\bm{x}\det_{1\leq i,j\leq N}\left[\text{Ai}(x_j-\beta_{n_i})\right]\,,\quad E^{(m)}_N=\frac{1}{3}N(N-1)\,.
}
and with no simple expression for the normalisation constant $A^{(m)}_N$.

\section{Inverse Laplace transform of $\widehat{\mathcal{S}}_N(\lambda, T)$}
\label{sec:ILT}

First of all, we start by noticing that  $\widehat{\mathcal{S}}_N(\lambda,T)=\widehat{\mathcal{Q}}_N(s)$ with $s=\lambda T^{3/2}$. This, as  pointed out  already in \cite{Majumdar2005}, implies a scaling law of the form $\mathcal{S}_N(A,T)=T^{-3/2}\mathcal{Q}_N(AT^{-3/2})$ so that $\widehat{\mathcal{Q}}_N(s)=\mathcal{L}[\mathcal{Q}_N(x)]$ with $x=AT^{-3/2}$. This applies to both cases of reunions and meanders. Notice that, even though this scaling appears mathematically, it can be easily derived by simple dimensional analysis. Indeed, as  the dimensions of the diffusion constant are $[D]=L^2 T^{-1}$ and the dimensions of area in our problem are $[A]=LT$ (Nb. here our $T$ refers to the time dimension, not to be confused with our final time $T$) then we have that
\beeqn{
\mathcal{S}_N(A,T)=\frac{1}{\sqrt{D T^3}}\mathcal{Q}_N\left(\frac{A}{\sqrt{DT^3}}\right)\,,
}
as we have found\footnote{For simplicity we have set $D=\frac{1}{2}$. This can be thought as equating dimensions of length squared with time, which implies that the dimensions of area are $[A]=T^{3/2}$ }. Secondly,  by looking at the exponents $E_N^{(r)}$ and $E_{N}^{(m)}$ (see either at table \ref{table2} or the expressions for the exponents in eqs. \eqref{eq:ltpdf1} and  \eqref{eq:ltpdf2}),  it is clear that we must perform the inverse Laplace transform  $\mathcal{L}^{-1}[s^{\delta}e^{s^{2/3}}]$ with $\delta$ either an integer or one-third of an integer.\\
To this end, we start from the result
\beeqn{
\int_0^{\infty}F(x)e^{-sx}dx=e^{-s^{2/3}}\,,\quad F(x)=2^{1/3}J(2^{1/3}x^{-2/3})x^{-5/3}
\,,
}
with
\beeqn{
J(x)=\frac{2^{2/3}x}{3^{3/2}\sqrt{\pi}}U(1/6,4/3,2x^3/27)e^{-\frac{2x^3}{27}}\,,
}
and with $U(a,b,x)$ the confluent hypergeometric function. After doing the rescalling of $s^{2/3}\to 2^{-1/3}\gamma s^{2/3}$, and a change of variables, we obtain the following formula
\beeqn{
s^{n+2k/3} e^{-\gamma 2^{-1/3}s^{2/3}}=(-1)^{k} 2^{k/3}\int_0^{\infty}\left[\frac{\partial^{k+n} F(x,\gamma)}{\partial \gamma^k \partial x^n} \right]e^{ sx}dx\,,\quad \quad F(x,\gamma)\equiv \gamma J(\gamma x^{-2/3} ) x^{-5/3}\,,
}
or alternatively
\beeq{
\mathcal{L}^{-1}[s^{n+2k/3} e^{-\gamma 2^{-1/3}s^{2/3}}]=(-1)^{k} 2^{k/3}\frac{\partial^{k+n}}{\partial \gamma^k \partial x^n} F(x,\gamma)\,.
\label{eq:alter}
}
Here we have used the property
\beeqn{
\mathcal{L}[F^{(n)}(x)]=s^{n}\widehat{F}(s)-\sum_{k=0}^{n-1}s^{k}F^{(n-1-k)}(0^{+})\,,
}
with notation $F^{(n)}(x)=d^{n} F(x)/dx^{n}$, and the fact that any derivative of $F(x,\gamma)$ with respect to $x$ at $x\to0^{+}$ is zero.\\
To get an idea of the order of the derivatives involved in the function $F(x,\gamma)$, one can   see how the exponents $E^{(r)}_N$ and $E_{N}^{(m)}$ vary with $N$  and choose a value of the pair $(n,k)$ which gives such an exponent. This choice is not necessarily unique and we show one possible choice in Table \ref{table2}.  The choice made is such that we only need derivatives with respect to $\gamma$ up to second order.
\begin{table}[t]
\begin{center}
\begin{tabular}{ |c||c|c|c||c|c|c|||c|c|c||c|c|c| }\hline
\multicolumn{1}{c}{$\#$ of Particles}& \multicolumn{6}{c}{Absorbing Boundary Conditions}& \multicolumn{6}{c}{Reflecting Boundary Conditions} \\
\hline
$N$&$E^{(r)}_N$&$n$&$k$&$E^{(m)}_N$&$n$&$k$&$E^{(r)}_N$&$n$&$k$&$E^{(m)}_N$&$n$&$k$\\\hline
1 & 1 & 1&0&$\frac{1}{3}$&-&-&$\frac{1}{3}$&-&-&0&0&0 \\
 2 & $\frac{10}{3}$ &2&2&$\frac{4}{3}$&0&2&2&2&0&$\frac{2}{3}$&0&1 \\
 3 & 7 & 7&0&3&3&0&5&5&0&2&2&0\\
 4 & 12 & 12&0&$\frac{16}{3}$&4&2&$\frac{28}{3}$&8&2&4&4&0 \\
 5 & $\frac{55}{3}$ &17 &2&$\frac{25}{3}$&7&2&15&15&0&$\frac{20}{3}$&6&1 \\
 6 & 26 & 26&0&12&12&0&22&22&0&10&10&0 \\
 7 & 35 & 35&0&$\frac{49}{3}$&15&2&$\frac{91}{3}$&29&2&14&14&0 \\
 8 & $\frac{136}{3}$ &44&2&$\frac{64}{3}$&20&2&40&40&0&$\frac{56}{3}$&18&1 \\
 9 & 57 & 57&0&27&27&0&51&51&0&24&24&0 \\
 10 & 70 & 70&0&$\frac{100}{3}$&32&2&$\frac{190}{3}$&62&2&30&30&0 \\\hline
\end{tabular}
\caption{Exponents $E^{(r)}_N$ and $E_{N}^{(m)}$ as a function of $N$ for both absorbing and reflecting boundary conditions. In the columns $n$ and $k$, we show the number of derivatives involved for the function $F(x,\gamma)$.  Recall that the expressions for the exponents are $E^{(r)}_N=N(2N+1)/3$ and $E_{N}^{(m)}=N^2/3$ for ABCs and  $E^{(r)}_N=N(2N-1)/3$ and $E_{N}^{(m)}=N(N-1)/3$ for RBCs.}
\label{table2}
\end{center}
\end{table}
The only case for which we are unable to use this prescription directly is for the case $N=1$ for meanders, as this will imply the exponent $E^{(m)}_1=1/3$. A way around this is to consider in this case the pair $(0,2)$ which will give the derivative of the PDF, instead.\\
This being settled,  we proceed to find a simple expression for any order derivative of the function $F(x,\gamma)$. Starting from
\beeqn{
F(x,\gamma)=\frac{\sqrt{3}}{x\sqrt{\pi}}u^{2/3}(x,\gamma)e^{-u(x,\gamma)}U\left(\frac{1}{6},\frac{4}{3},u(x,\gamma)\right)\,,\quad u(x,\gamma)=\frac{2\gamma^3}{27x^2}\,,
}
and using properties of the hypergeometric confluent function we arrive at the following result (see appendix \ref{app:dF})
\beeq{
\frac{\partial^{k+n} F(x,\gamma)}{\partial \gamma^k\partial x^n}=\frac{\sqrt{3}}{x^{n+1}\sqrt{\pi}\gamma^{k}}u^{2/3}(x,\gamma)e^{-u(x,\gamma)}\sum_{\ell=0}^{n}\sum_{s=0}^{k}C^{(n)}_{\ell} D^{(k)}_{s}(\ell)U\left(\frac{1}{6}-\ell-s,\frac{4}{3},u(x,\gamma)\right)\,,
\label{eq:res2}
}
where the set of coefficients $\{C^{(n)}_\ell\}$ and $\{D^{(k)}_{s}(\ell)\}$ are given by
\beeqn{
C^{(n)}_\ell&=\frac{n!}{2^{n-2\ell} (n-\ell)!(2\ell-n)!}\,,\quad \ell=0,\ldots,n\,,\\
D_0^{(0)}(\ell)&=1\,,\quad\quad D_{0}^{(1)}(\ell)=-\frac{3}{2}\left(1+2\ell\right)\,,\quad\quad D_{1}^{(1)}(\ell)=-3\,\\
D_0^{(2)}(\ell)&=\frac{3}{4} (1 + 2\ell) (5 + 6\ell)\,,\quad\quad D_1^{(2)}(\ell)=3(7+6\ell)\,,\quad\quad D_2^{(2)}(\ell)=9\,.
}
Notice that for the set of coefficients $\{D^{(k)}_{s}(\ell)\}$ we have already taken into account the fact that we only need derivatives with respect to $\gamma$  up to second order.\\
With the help of eqs. \eqref{eq:alter} and \eqref{eq:res2}, we can now perform the inverse Laplace transform of $\widehat{\mathcal{Q}}^{(r)}_N(s)$ and $\widehat{\mathcal{Q}}^{(m)}_N(s)$, obtaining the following generalised Airy distributions for absorbing boundary conditions:
\begin{eqnarray}
\mathcal{Q}^{(r)}_N(x)&=&\sum_{n_1,\ldots,n_N} \Delta^2_N(\alpha_{n_1},\dots \alpha_{n_N})\mathcal{I}^{(n,k)}_{N,(r)}\left(x,\sum_{i=1}^N\alpha_{n_i}\right)\,,\label{eq:gad}\\
\mathcal{Q}^{(m)}_N(x)&=&\sum_{n_1,\ldots,n_N}B^{(A)}_N(\alpha_{n_1},\dots \alpha_{n_N}) \Delta_N(\alpha_{n_1},\dots \alpha_{n_N})\mathcal{I}^{(n,k)}_{N,(m)}\left(x,\sum_{i=1}^N\alpha_{n_i}\right)\,.
\label{eq:gad2}
\end{eqnarray}
For reflecting boundary conditions we have instead
\begin{eqnarray}
\mathcal{Q}^{(r)}_N(x)&=&\sum_{n_1,\ldots,n_N} \frac{\Delta^2_N(\beta_{n_1},\dots \beta_{n_N})}{\beta_{n_1}\cdots\beta_{n_N}}\mathcal{I}^{(n,k)}_{N,(r)}\left(x,\sum_{i=1}^N\beta_{n_i}\right)\,,\label{eq:gadR}\\
\mathcal{Q}^{(m)}_N(x)&=&\sum_{n_1,\ldots,n_N}B^{(R)}_N(\beta_{n_1},\dots \beta_{n_N}) \frac{\Delta_N(\beta_{n_1},\dots \beta_{n_N})}{\beta_{n_1}\cdots\beta_{n_N}}\mathcal{I}^{(n,k)}_{N,(m)}\left(x,\sum_{i=1}^N\beta_{n_i}\right)\,.
\label{eq:gad2R}
\end{eqnarray}

In both cases we have defined
\beeqn{
\mathcal{I}^{(n,k)}_{N,(s)}(x,\gamma)=\frac{(-1)^{k} 2^{k/3}}{A^{(s)}_{N}}\frac{\sqrt{3}}{x^{n+1}\sqrt{\pi}\gamma^{k}}u^{2/3}(x,\gamma)e^{-u(x,\gamma)}\sum_{\ell=0}^{n}\sum_{s=0}^{k}C^{(n)}_{\ell} D^{(k)}_{s}(\ell)U\left(\frac{1}{6}-\ell-s,\frac{4}{3},u(x,\gamma)\right)\,.
}
with $s\in\{r(eunion),m(eanders)\}$. In the expression of $\mathcal{I}^{(n,k)}_{N,(s)}(x,\gamma)$  the pair of indices $(n,k)$ must be chosen according to the values related to the exponents $E^{(r)}_N$ and $E_{N}^{(m)}$ appearing in table \ref{table2}. Notice, in particular, that for $N=1$ and $(n,k)=(1,0)$ we have that $A^{(r)}_1=1/\sqrt{2\pi}$, $\Delta_N=1$, $C_0^{(1)}=0$, $C_1^{(1)}=2$. Thus $\mathcal{Q}^{(r)}_1(x)$  we recover to the so-called Airy distribution, \textit{viz}
\beeqn{
\mathcal{Q}^{(r)}_1(x)&=\sum_{n=1}^\infty \mathcal{I}^{(1,0)}_{1,(r)}\left(x,\alpha_{n}\right)\,,\quad\quad \mathcal{I}^{(1,0)}_{1,(r)}(x,\gamma)=\frac{2\sqrt{6}}{x^2}\left(\frac{2\gamma^3}{27x^2}\right)^{2/3}e^{-\frac{2\gamma^3}{27x^2}}U\left(-\frac{5}{6},\frac{4}{3},\frac{2\gamma^3}{27x^2}\right)\,.
}

\section{Moments}
\label{sec:M}
In this section we discuss the derivation of the moments. We consider negative powered-moments. Let us start by fixing some notation. Let us denote $M_n$ the $n$-th moment for a PDF $F(x)$ with $x\in\mathbb{R}^{+}$ $A_n=\int_0^\infty dx x^n F(x)$. Suppose next that the Laplace transform of a function $F(x)$ is  $\widehat{F}(s)=s^{E}e^{-as^{2/3}}$, with $E$ and $a$ two constants. Then one can show that (see appendix \ref{app:moments})
\beeqn{
A_{-\nu}=\frac{3}{2}a^{-\frac{3}{2}(\nu+E)}\frac{\Gamma\left(\frac{3}{2}(\nu+E)\right)}{\Gamma(\nu)}\,.
\label{den1}
}
In our case let us denote  the moments as $M^{(a)}_{N,n}=\int_0^\infty dx \mathcal{Q}^{(a)}_{N}(x)x^n$ with $a\in\{r,m\}$. Using the previous results, we obtain the following formulas for the negative moments of the PDF for reunions and meanders with ABCs
\beeqn{
M^{(r)}_{N,-\nu}&=2^{\frac{\nu+E^{(r)}_N}{2}-1}\frac{3\Gamma\left(\frac{3}{2}(\nu+E^{(r)}_N)\right)}{A_N^{(r)}\Gamma(\nu)}\sum_{n_1,\ldots,n_N}\Delta^2_N(\alpha_{n_1},\ldots,\alpha_{n_N})\left(\sum_{i=1}^N\alpha_{n_i}\right)^{-\frac{3}{2}(\nu+E^{(r)}_N)}\\
M^{(m)}_{N,-\nu}&=2^{\frac{\nu+E^{(m)}_N}{2}-1}\frac{3\Gamma\left(\frac{3}{2}(\nu+E^{(m)}_N)\right)}{A_N^{(m)}\Gamma(\nu)}\sum_{n_1,\ldots,n_N}\Delta_N(\alpha_{n_1},\ldots,\alpha_{n_N})B_N(\alpha_{n_1},\ldots,\alpha_{n_N})\left(\sum_{i=1}^N\alpha_{n_i}\right)^{-\frac{3}{2}(\nu+E^{(m)}_N)}
}
Similar expressions can be found for RBCs.

\section{Monte Carlo Simulations}
\label{sec:MCS}
To check the correctness of our analytical findings, we have performed Monte Carlo simulations exploiting the connection with Random Matrix Theory \cite{Katori2004,Kobayashi2008} to generate samples of non-colliding paths. Following the notation in \cite{Kobayashi2008} we first recall the definition of the Pauli matrices
\beeqn{
\sigma_1=\begin{pmatrix}
0&1\\
1&0
\end{pmatrix}\,,\quad\quad \sigma_2=\begin{pmatrix}
0&-i\\
i&0
\end{pmatrix}\,,\quad\quad \sigma_3=\begin{pmatrix}
1&0\\
0&-1
\end{pmatrix}\,,
}
while we denote the 2$\times 2$ identity matrix $I_2$ as  $\sigma_0$.  For meanders we define the following $2N\times 2N$ Hermitian matrices $\Xi_{T}^{C}(t)$ and $\Xi^D_T(t)$ corresponding to absorbing and reflecting boundary conditions
\beeqn{
\Xi^{C}_T(t)&=i a_T^{(0)}(t,O)\otimes \sigma_0+s^{(1)}(t)\otimes\sigma_1+s_T^{(2)}(t,O)\otimes\sigma_2+s^{(3)}(t)\otimes\sigma_3\,,\\
\Xi^{D}_T(t)&=i a_T^{(0)}(t,O)\otimes \sigma_0+i a_T^{(1)}(t,O)\otimes\sigma_1+i a^{(2)}(t)\otimes\sigma_2+s^{(3)}(t)\otimes\sigma_3\,,
}
respectively. Here the notation is a bit involved but it means the following: $(\cdots)_T(t,O)$ stands for bridges starting at the origin and finishing at the origin at time $T$, while $a^{(a)}$ and $s^{(a)}$ denote $N\times N$ antisymmetric and symmetric matrices, respectively, whose elements are either bridges or standard Brownian motions. Being more precise, if we denote as $b(t)$ a standard Brownian motion with $b(0)=0$, and with $B(t)$ a Brownian bridge with $B(0)=B(T)=0$ then
\beeqn{
s^{(a)}_{ij}(t)=\left\{
\begin{array}{ll}
b_{ij}^{(a)}(t)/\sqrt{2}& i<j\\
b_{ii}^{(a)}(t)& i=j\\
b_{ji}^{(a)}(t)/\sqrt{2}& j<i
\end{array}
\right.\,,\quad\quad a^{(a)}_{ij}(t)=\left\{
\begin{array}{ll}
\widetilde{b}_{ij}^{(a)}(t)/\sqrt{2}& i<j\\
0& i=j\\
-\widetilde{b}_{ji}^{(a)}(t)/\sqrt{2}& j<i
\end{array}
\right.
}
and
\beeqn{
s^{(a)}_{T,ij}(t,O)=\left\{
\begin{array}{ll}
B_{ij}^{(a)}(t)/\sqrt{2}& i<j\\
B_{ii}^{(a)}(t)& i=j\\
B_{ji}^{(a)}(t)/\sqrt{2}& j<i
\end{array}
\right.\,,\quad\quad a^{(a)}_{T,ij}(t,O)=\left\{
\begin{array}{ll}
\widetilde{B}_{ij}^{(a)}(t)/\sqrt{2}& i<j\\
0& i=j\\
-\widetilde{B}_{ji}^{(a)}(t)/\sqrt{2}& j<i
\end{array}
\right.\,.
}
Here the index $a=0,1,2,3$, simply states that we need to construct different matrices. Similarly, for the case of reunions we have instead the following two $2N\times 2N$ Hermitian matrices for absorbing and reflecting boundary conditions
\beeqn{
\Xi^{C}_T(t)&=i a_T^{(0)}(t,O)\otimes \sigma_0+s^{(1)}_T(t,O)\otimes\sigma_1+s_T^{(2)}(t,O)\otimes\sigma_2+s^{(3)}_T(t,O)\otimes\sigma_3\,,\\
\Xi^{D}_T(t)&=i a_T^{(0)}(t,O)\otimes \sigma_0+i a_T^{(1)}(t,O)\otimes\sigma_1+i a^{(2)}_T(t,O)\otimes\sigma_2+s^{(3)}_T(t,O)\otimes\sigma_3\,,
}
respectively.\\
Bridges are easily generated from a standard Brownian motion. In its discrete version if $R(k)$ is a standard random walk with $k\in\{0,\ldots,K\}$ with $R(k=0)=0$ then a discrete Brownian Bridge is given by $B(k)=R(k)-(k/K)R(k)$.
\begin{figure}
\includegraphics[width=3.8cm, height=3.5cm]{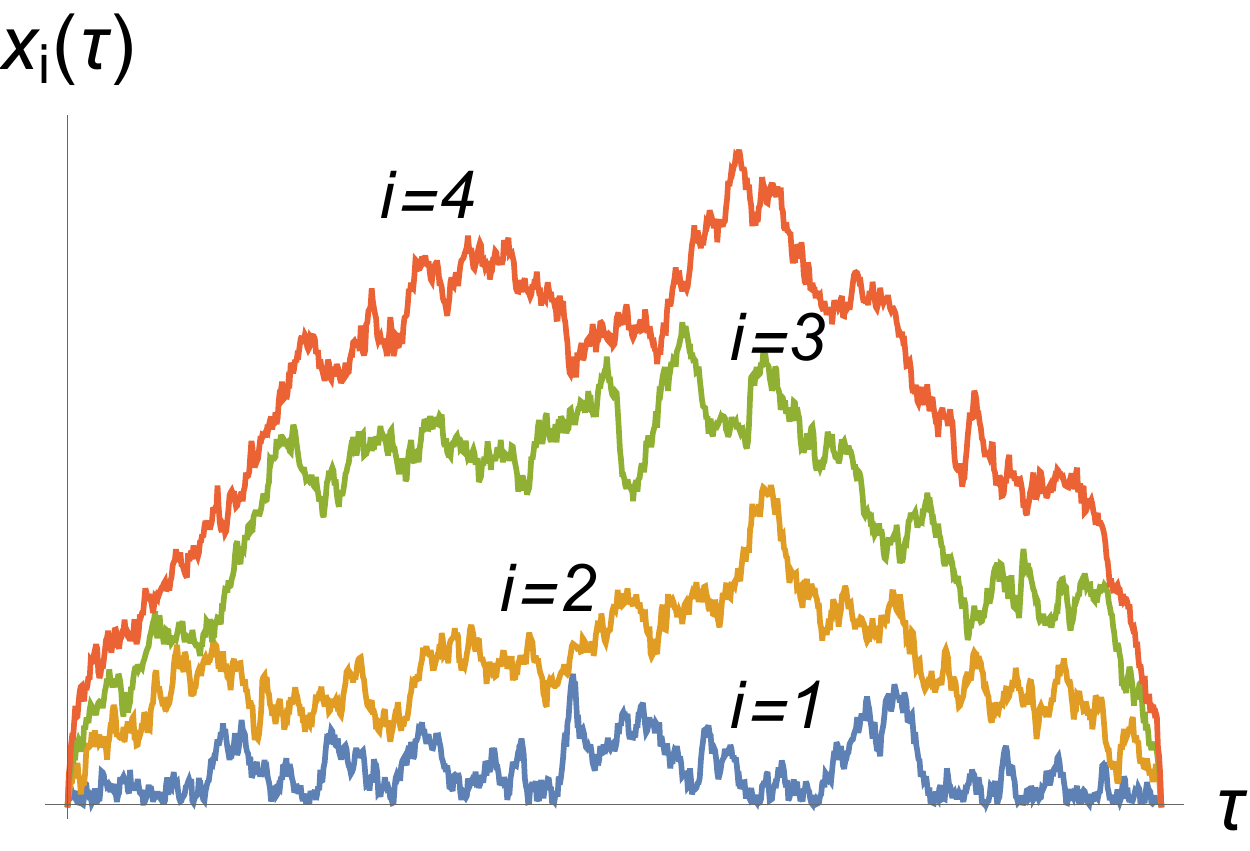}
\includegraphics[width=3.8cm, height=3.5cm]{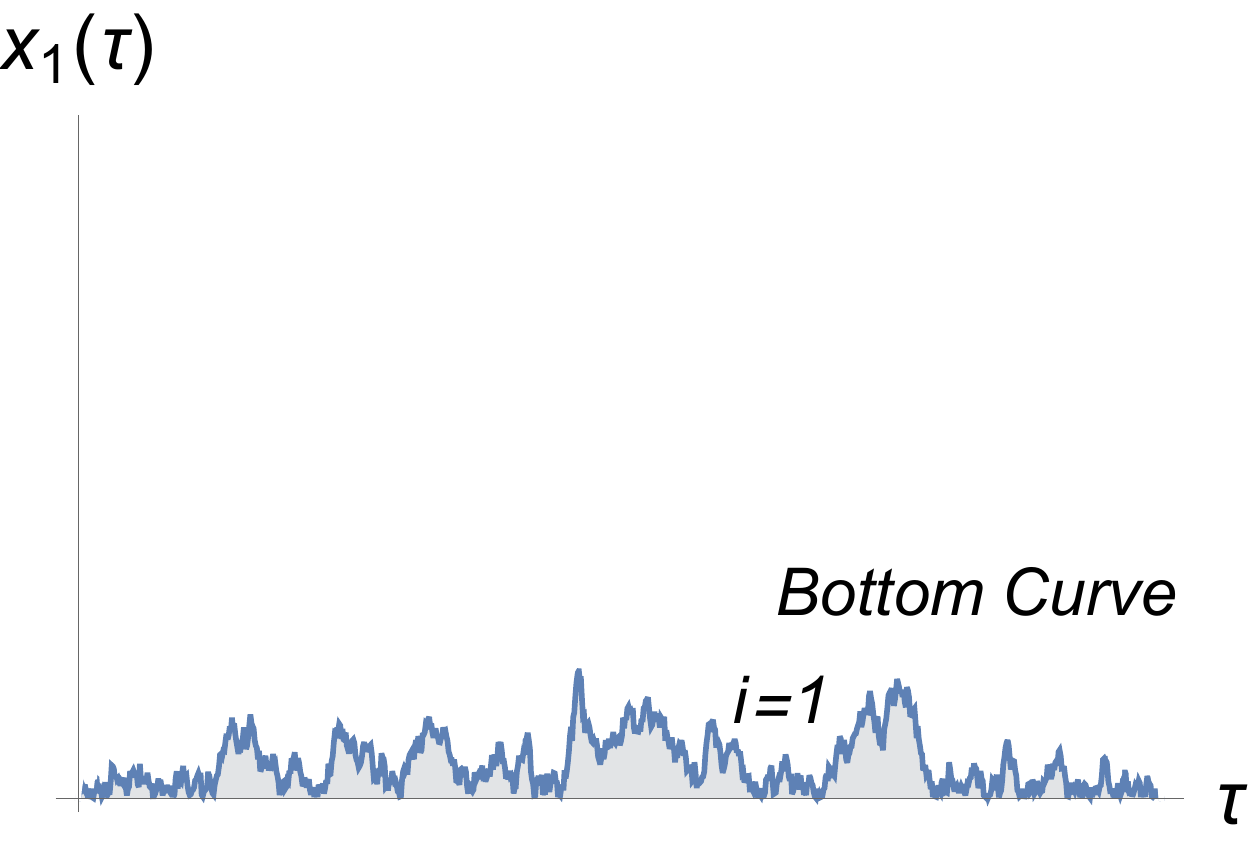}
\includegraphics[width=3.8cm, height=3.5cm]{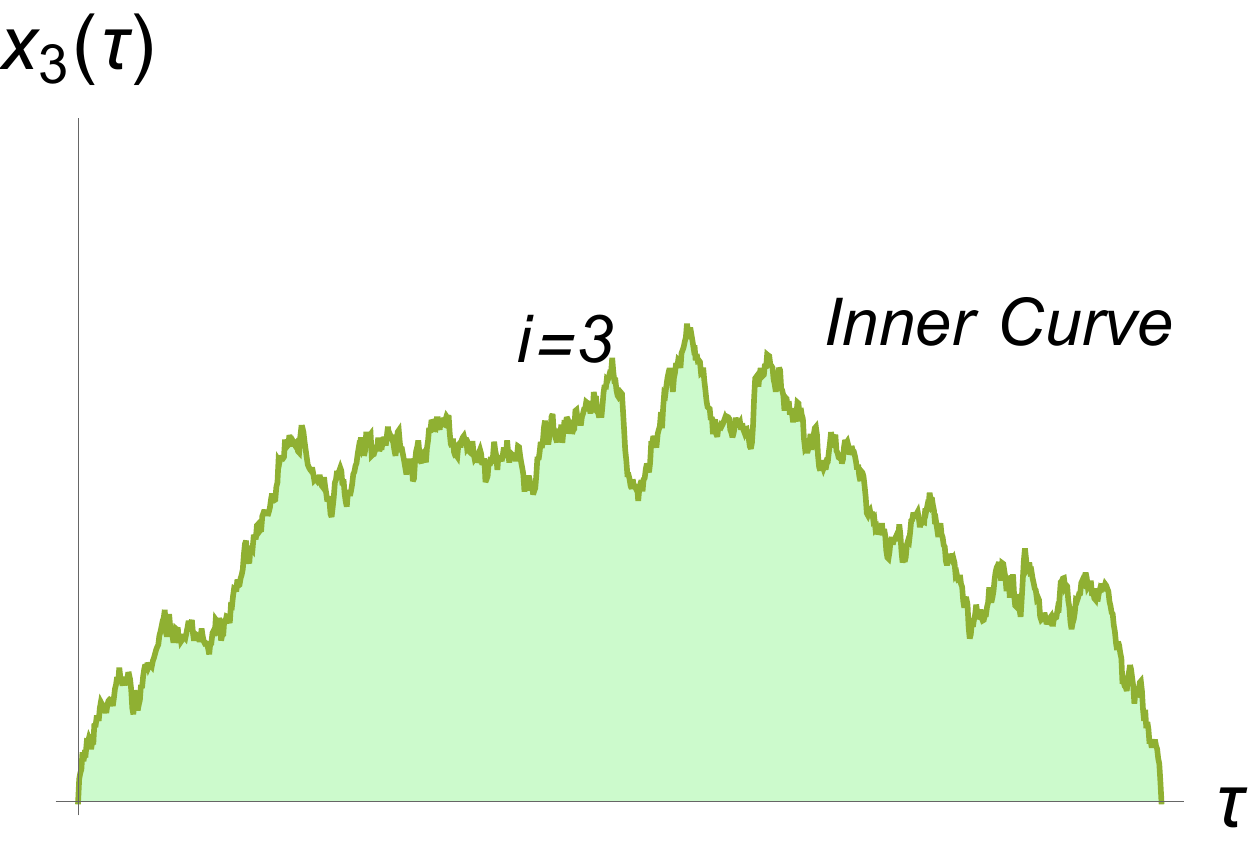}
\includegraphics[width=3.8cm, height=3.5cm]{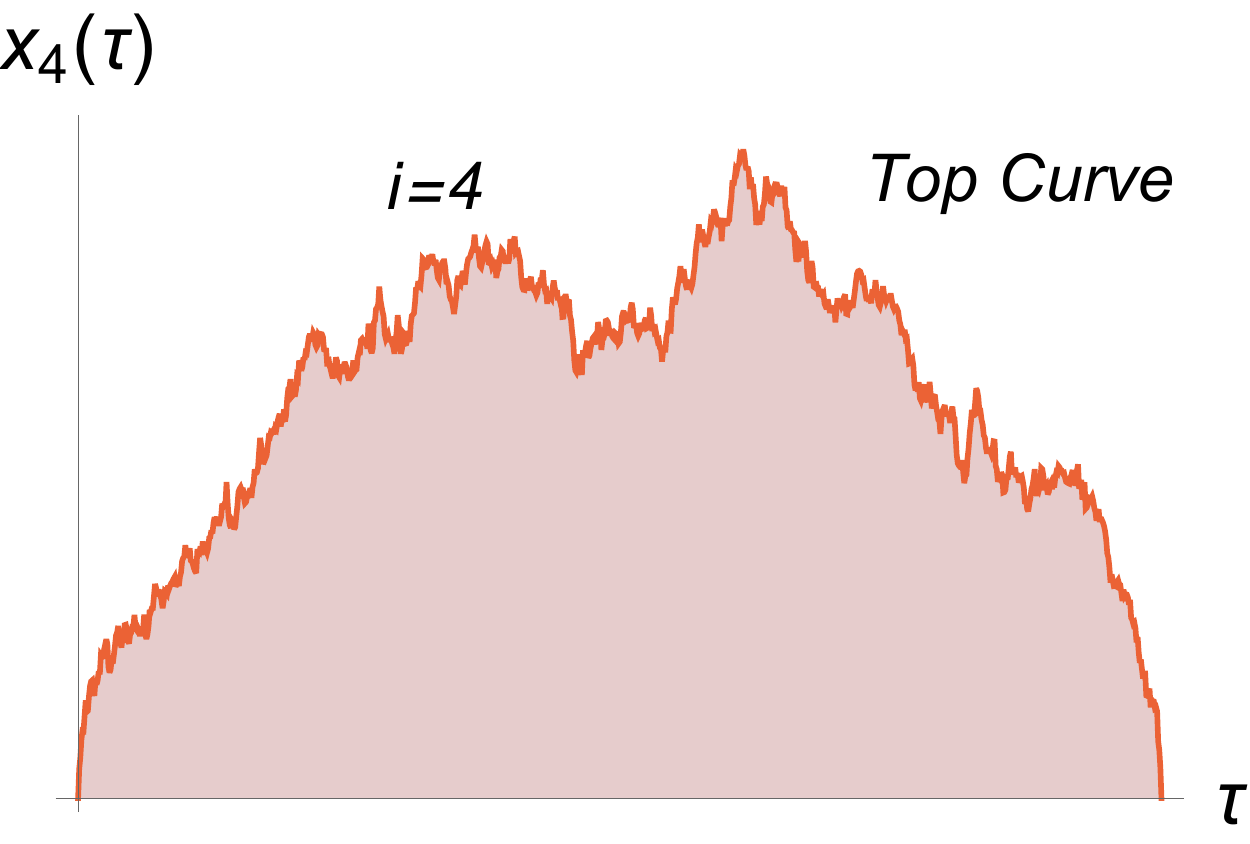}\\
\includegraphics[width=3.8cm, height=3.5cm]{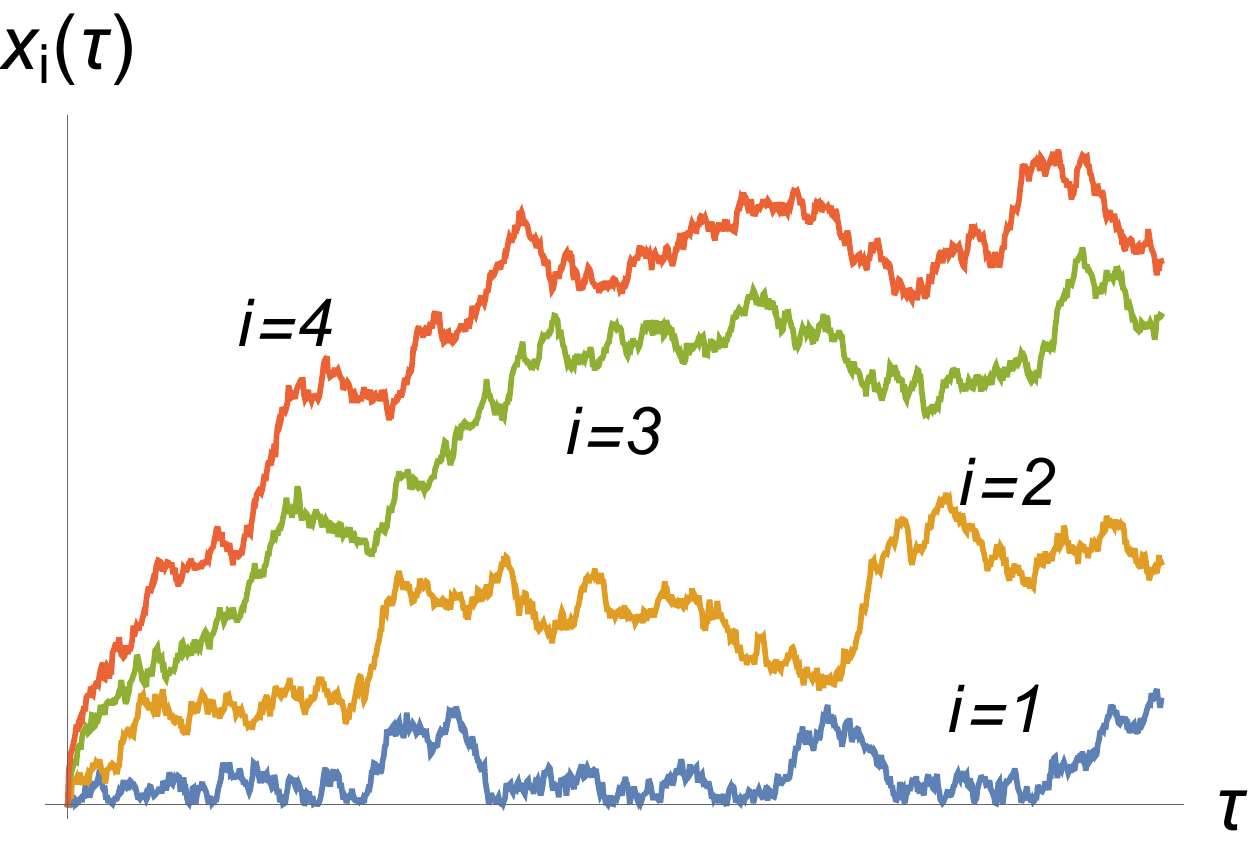}\includegraphics[width=3.8cm, height=3.5cm]{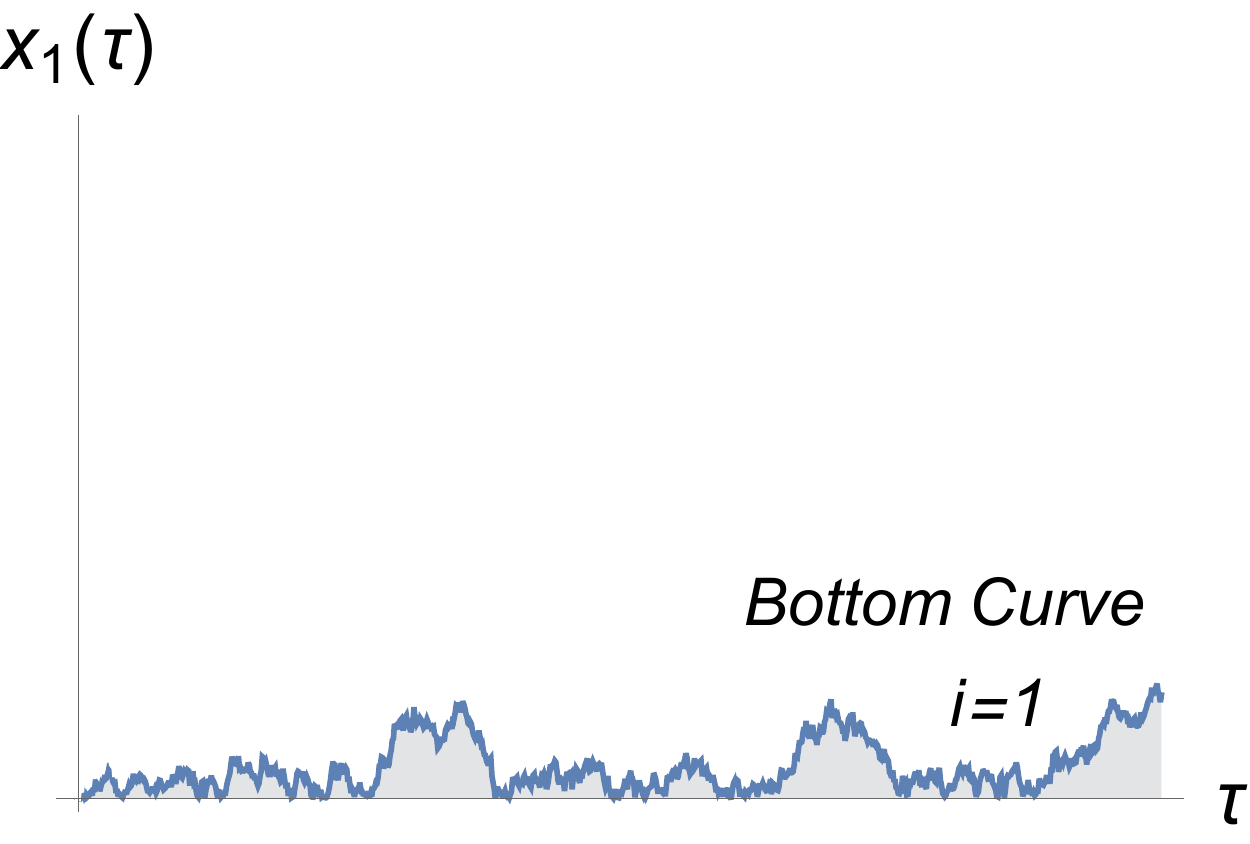}\includegraphics[width=3.8cm, height=3.5cm]{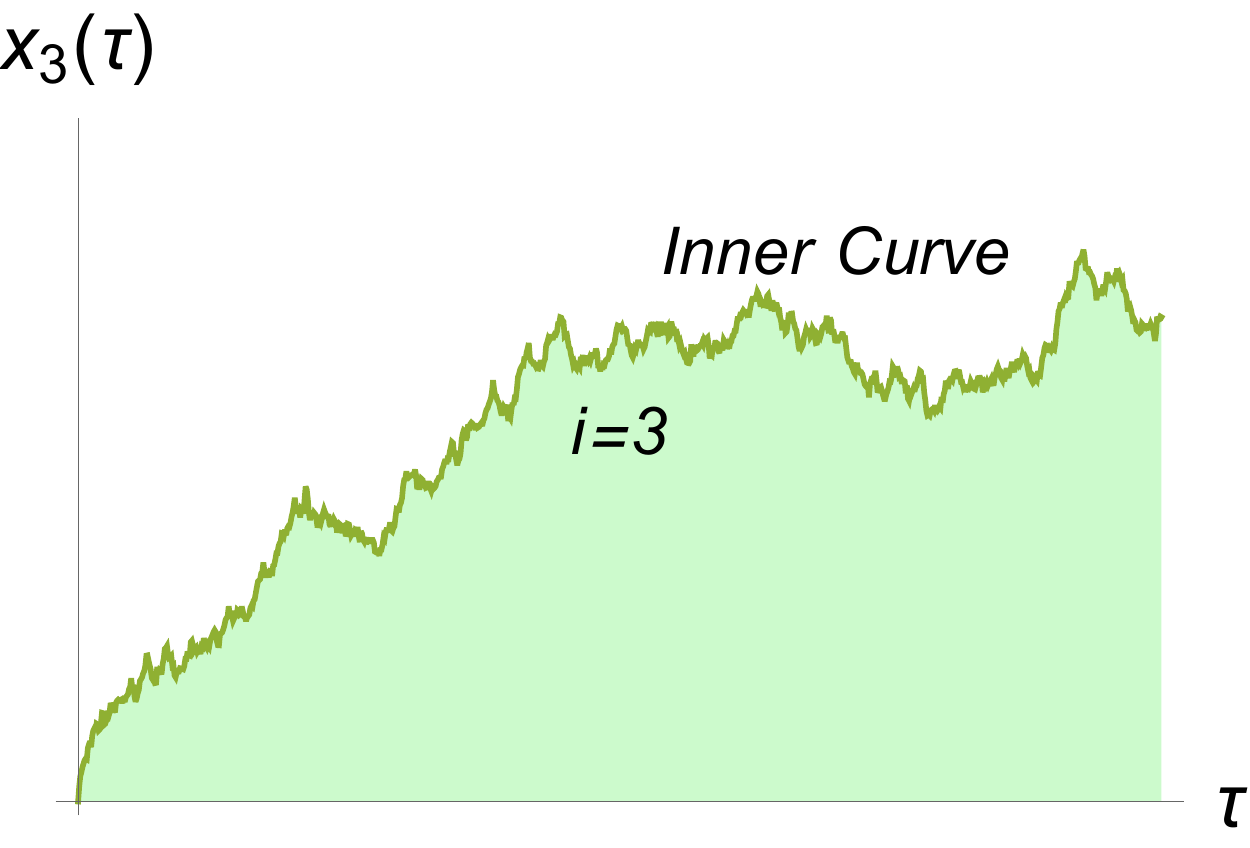}\includegraphics[width=3.8cm, height=3.5cm]{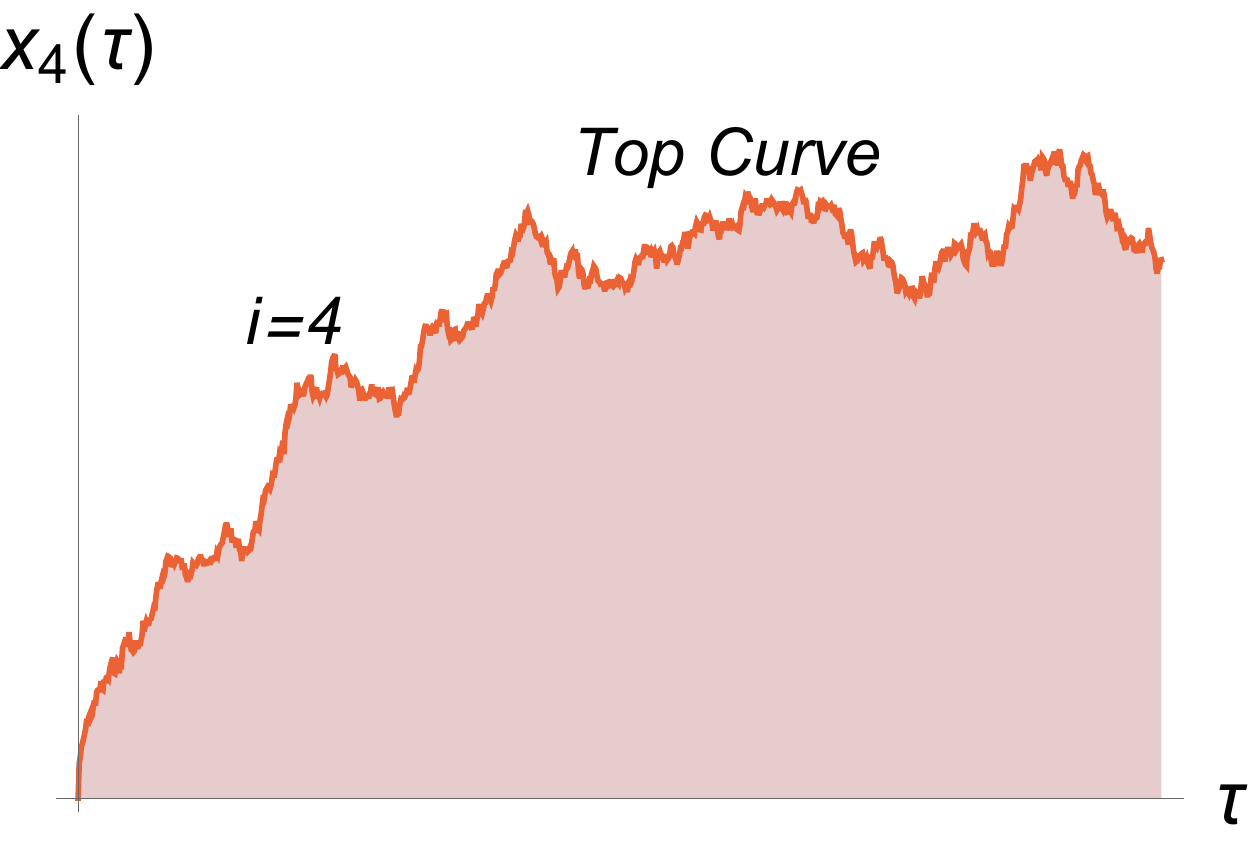}
\caption{Two instances corresponding to a reunion (top panel) and a meander (bottom panel) of $N$ Vicious paths with reflective boundary conditions at $x=0$, generated by the method explained in the text. We  also show the corresponding areas for the bottom, one of the inners, and the top curves.}
\label{fig2}
\end{figure}

\begin{figure}
\includegraphics[width=3.8cm, height=3.5cm]{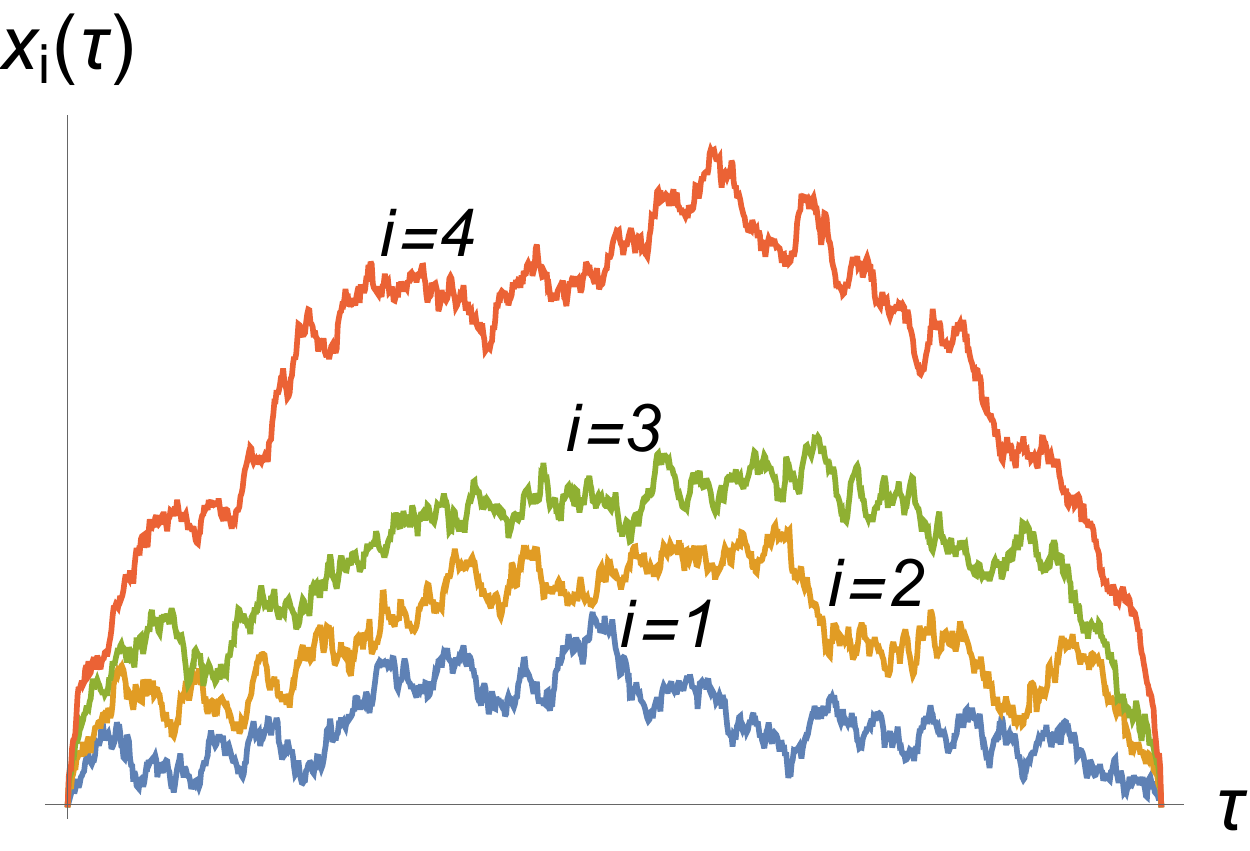}
\includegraphics[width=3.8cm, height=3.5cm]{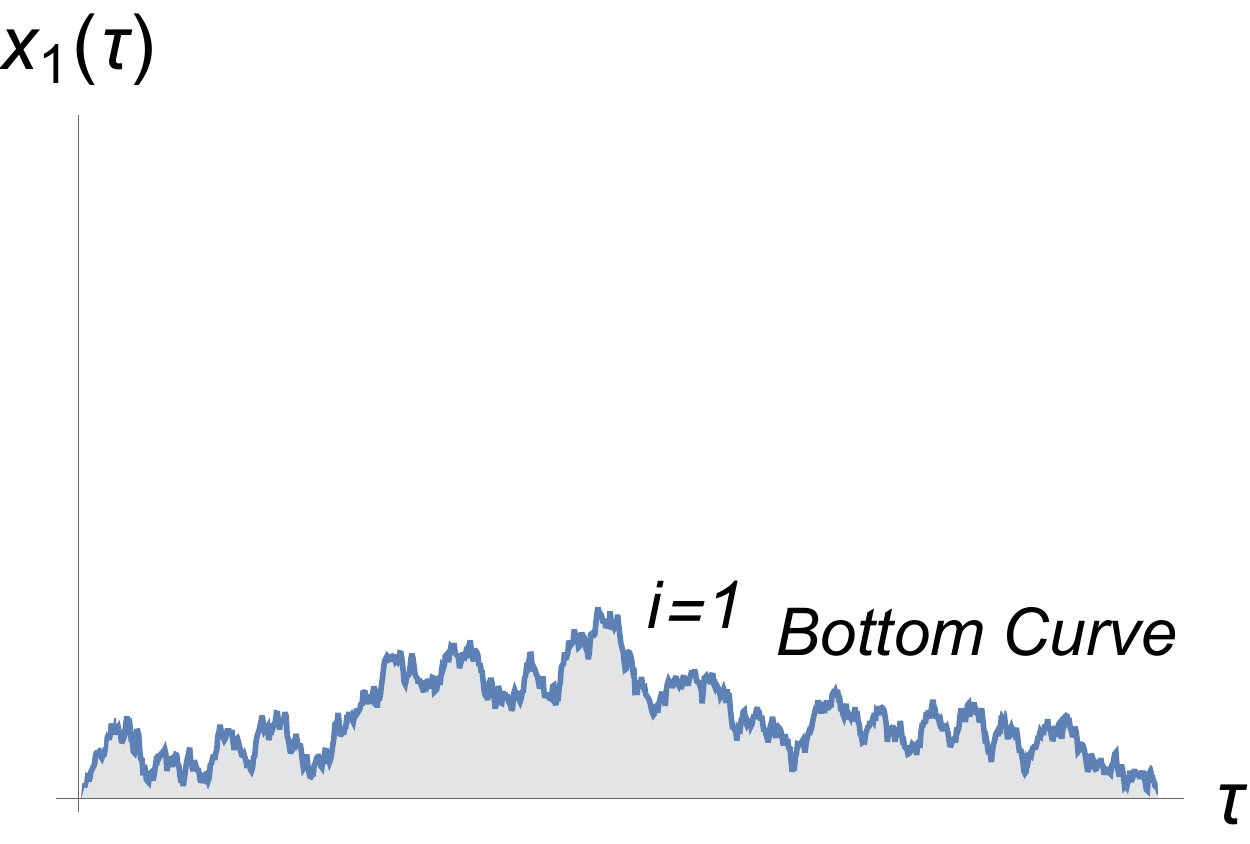}
\includegraphics[width=3.8cm, height=3.5cm]{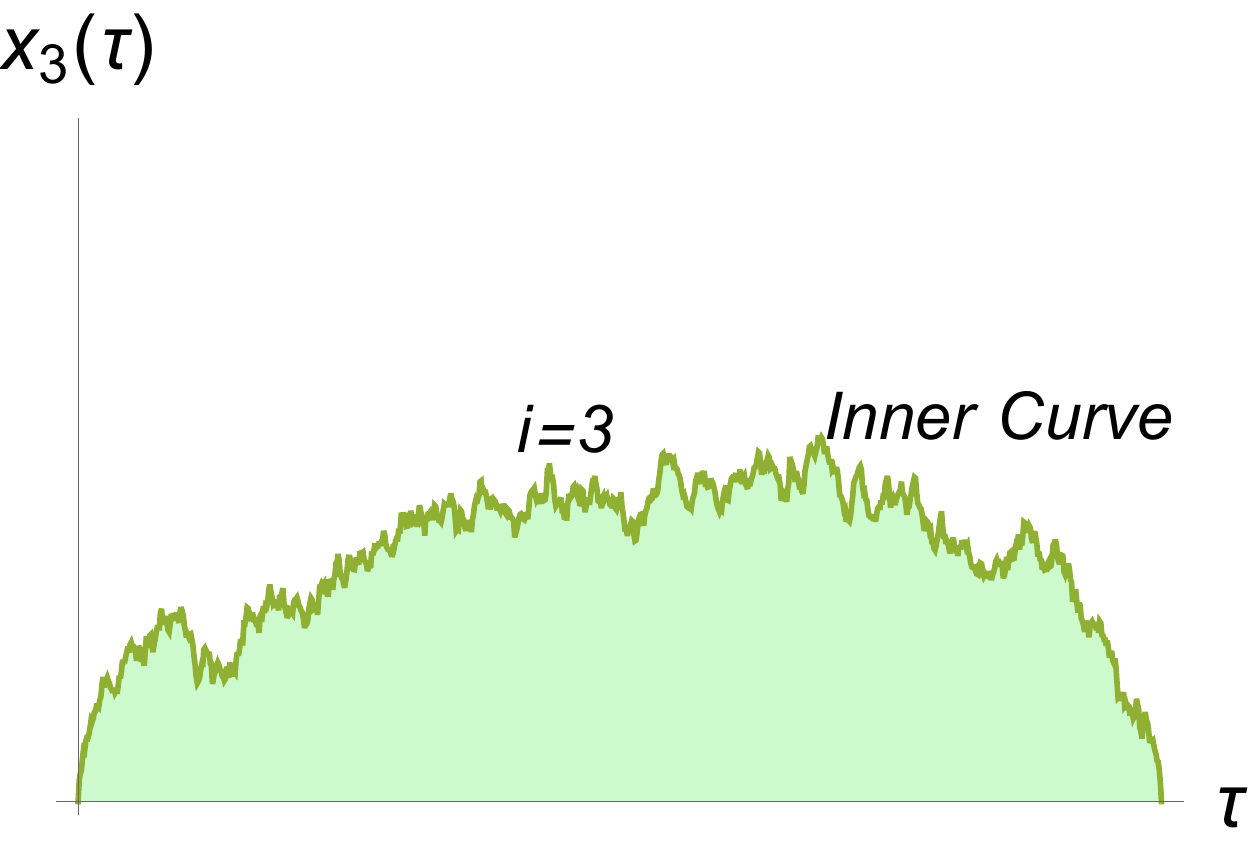}
\includegraphics[width=3.8cm, height=3.5cm]{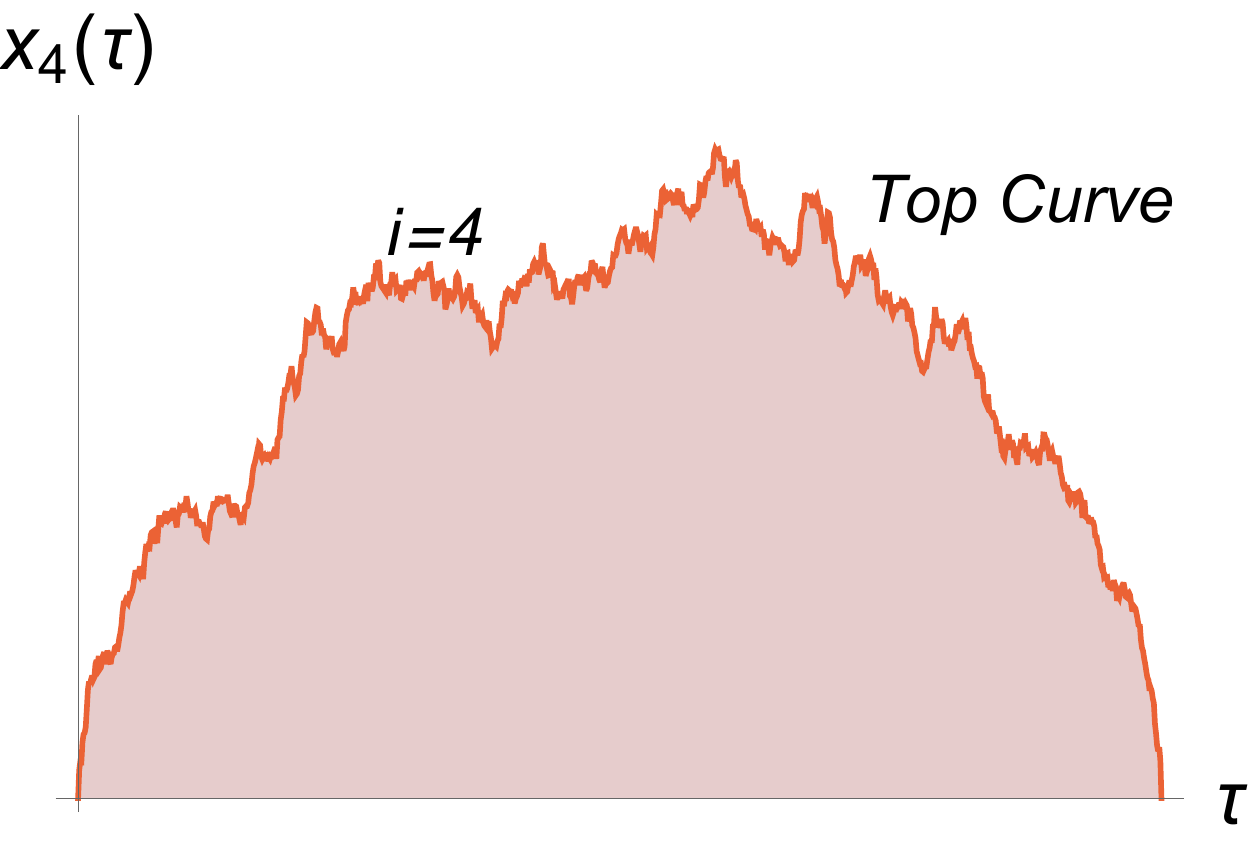}\\
\includegraphics[width=3.8cm, height=3.5cm]{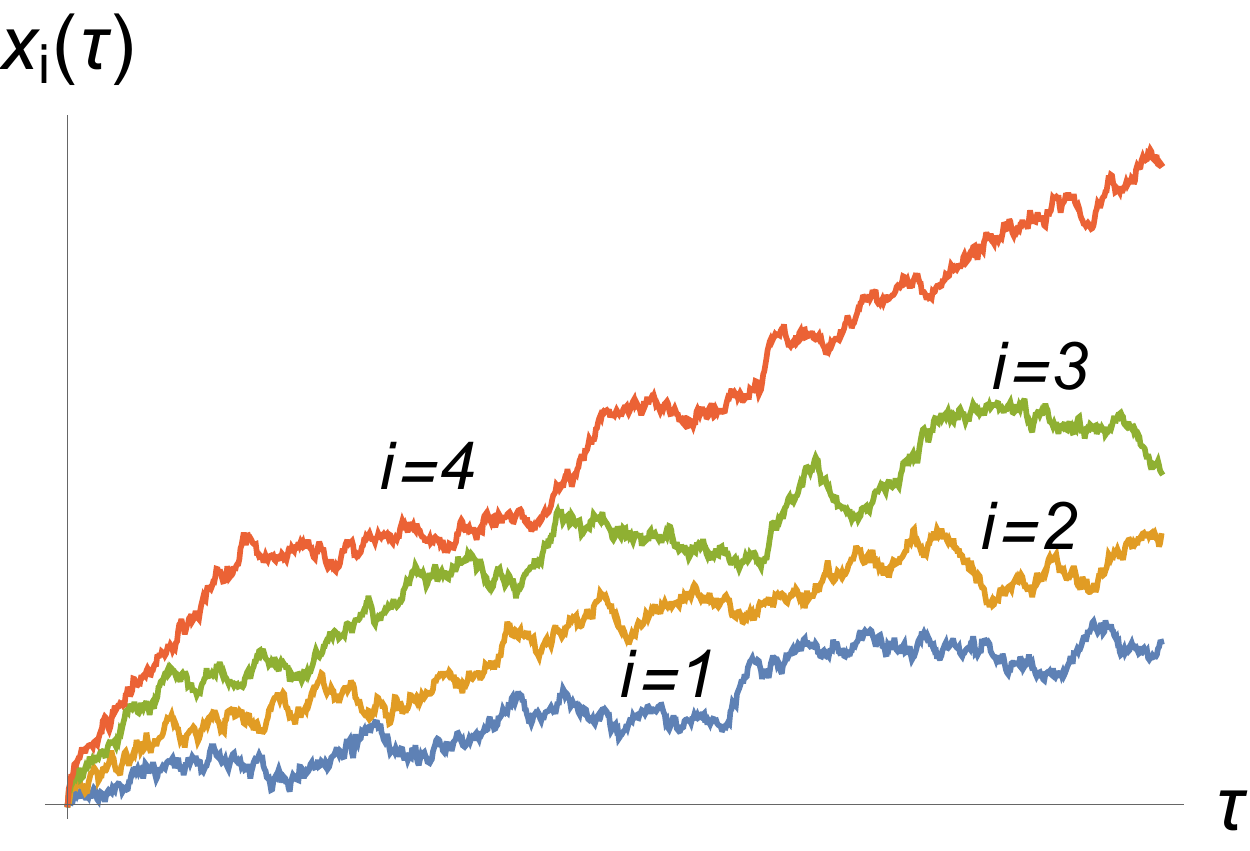}
\includegraphics[width=3.8cm, height=3.5cm]{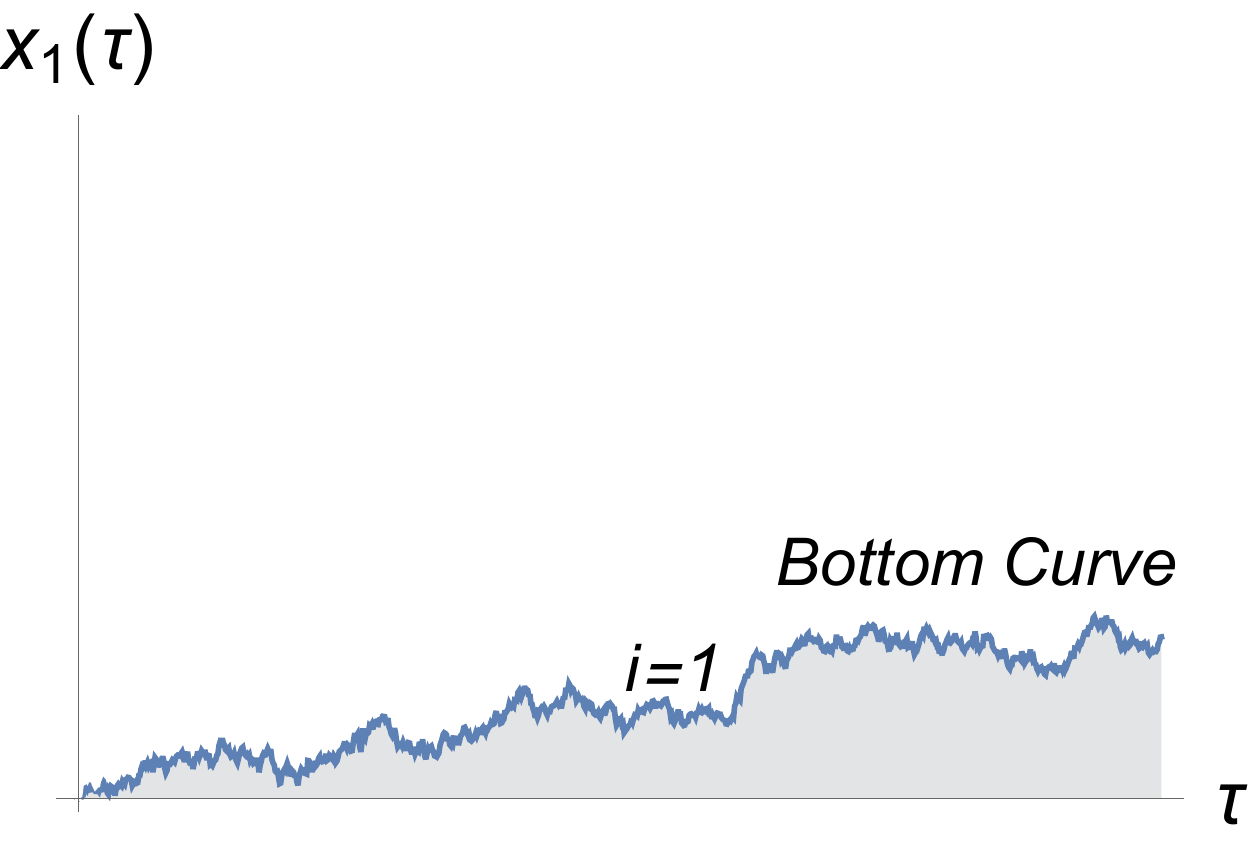}
\includegraphics[width=3.8cm, height=3.5cm]{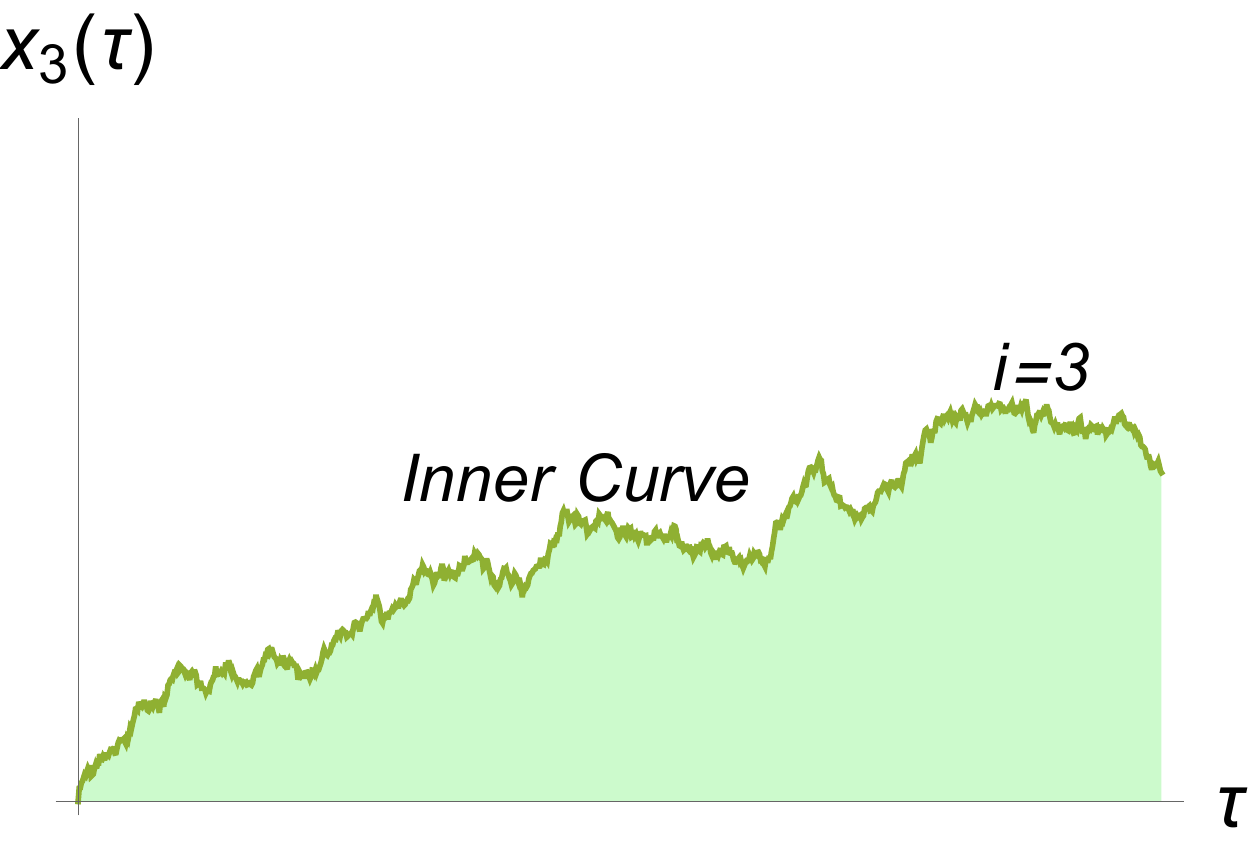}
\includegraphics[width=3.8cm, height=3.5cm]{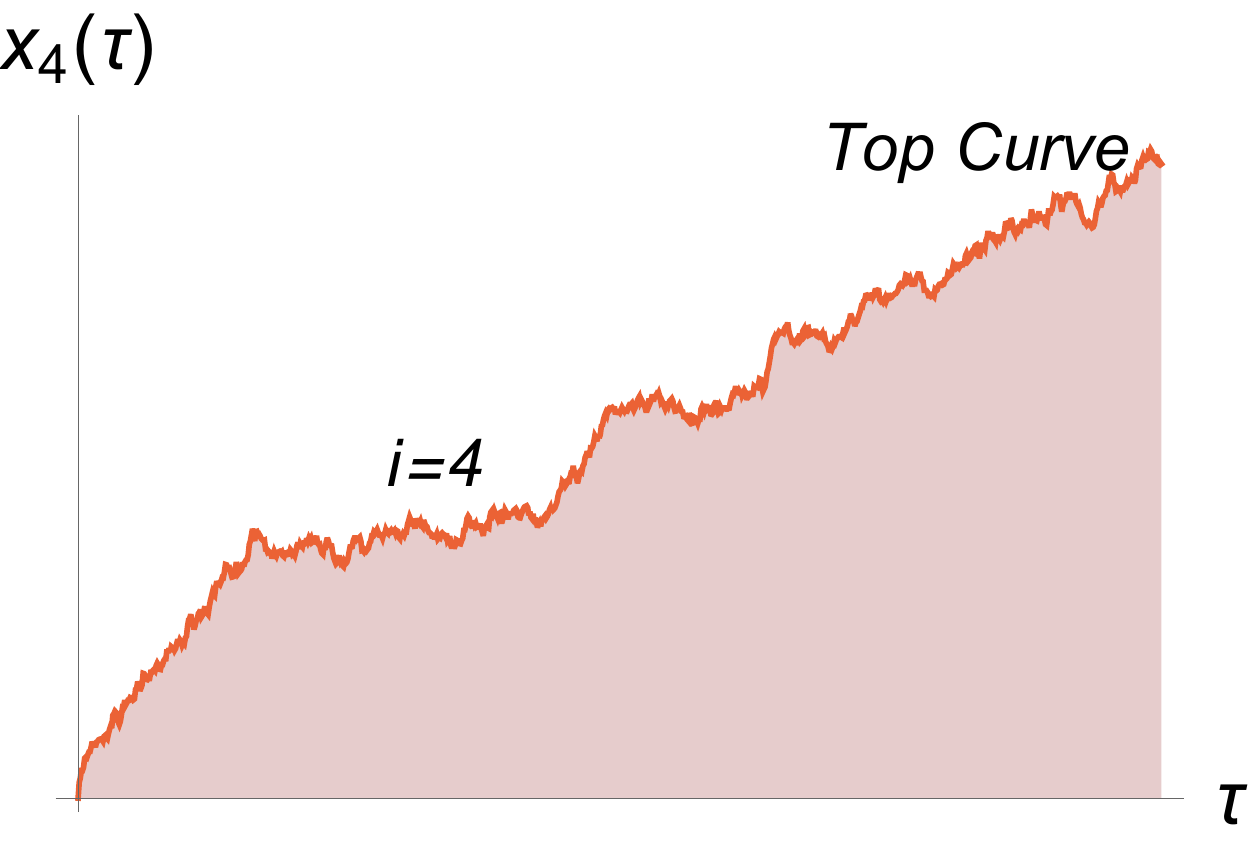}
\caption{Two instances corresponding to a reunion (top panel) and a meander (bottom panel) of $N$ Vicious paths with absorbing boundary conditions at $x=0$ and generated by the method explained in the text. We  also show the corresponding areas for the bottom, one of the inners, and the top curves.}
\label{fig3}
\end{figure}
\begin{figure}[t]
 \includegraphics[width=8cm, height=4cm]{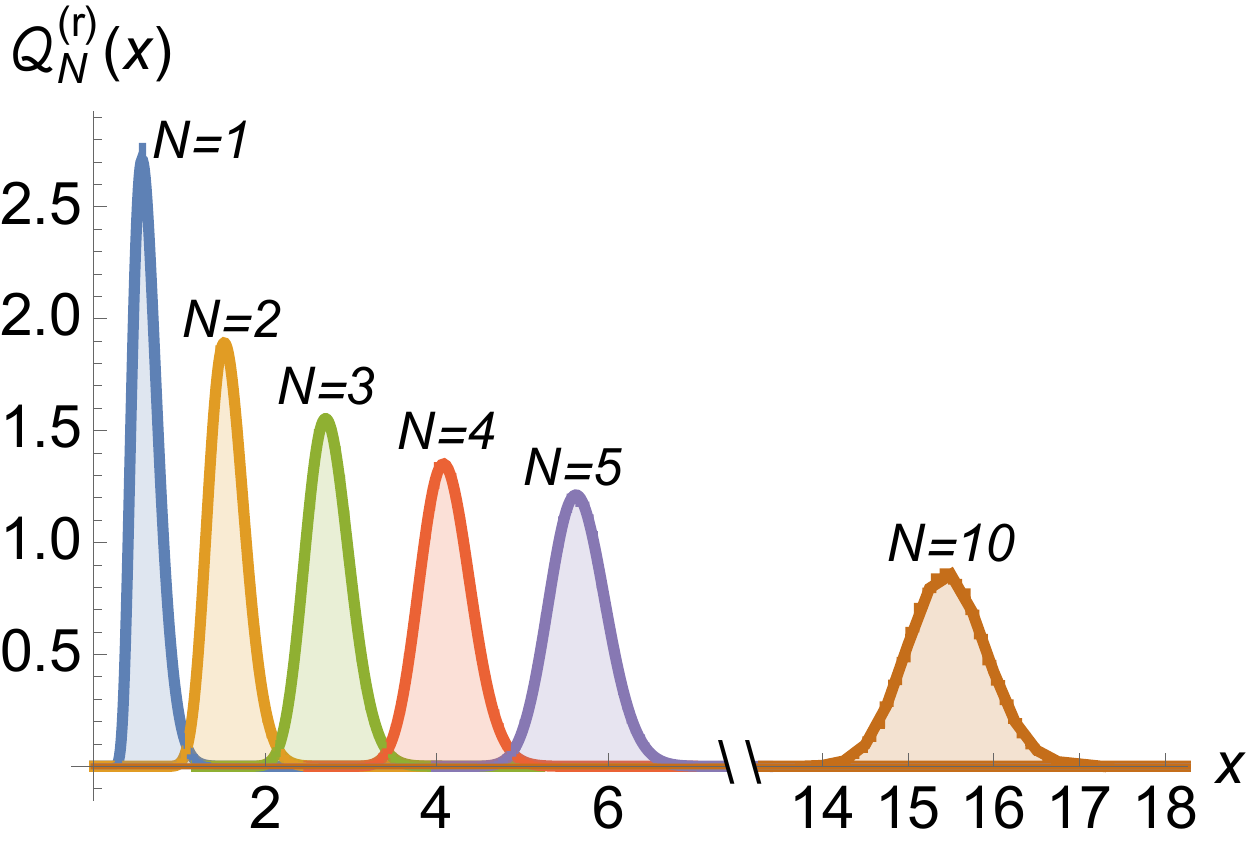}\includegraphics[width=8cm, height=4cm]{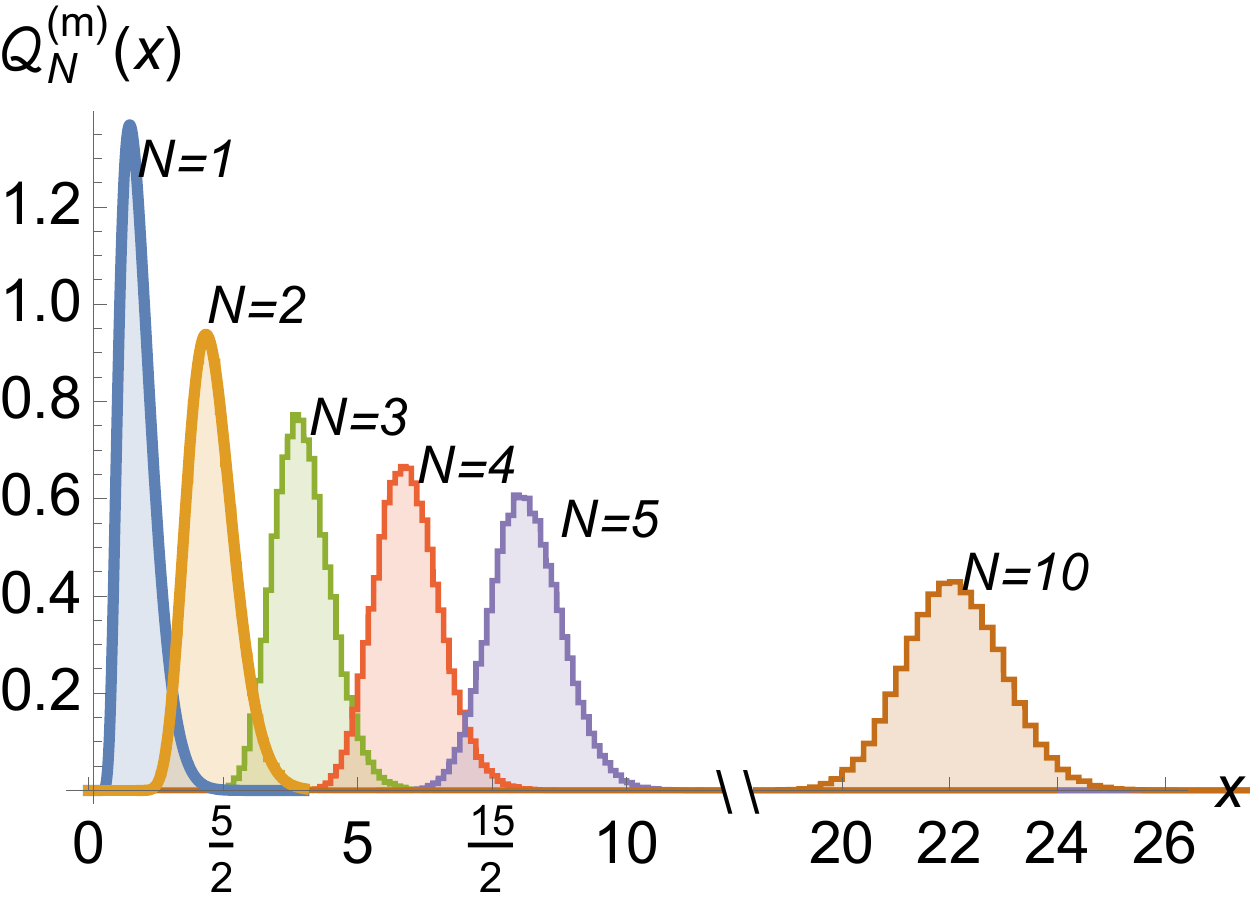}\\
 \includegraphics[width=8cm, height=4cm]{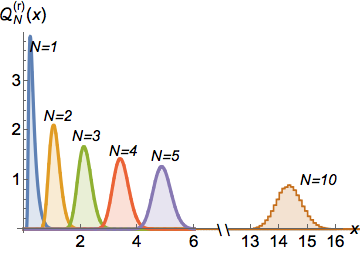}  \includegraphics[width=8cm, height=4cm]{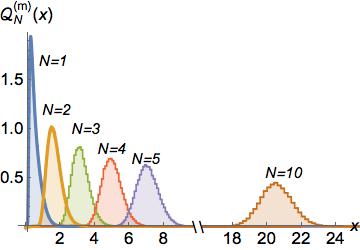}
\caption{Plots for $\mathcal{Q}^{(r,m)}_N(x)$  for $N=1,2,3,4$ and $10$. The top plots corresponds to absorbing boundary conditions, while the bottom plots are for reflecting boundary conditions. Thick solid lines the theoretical results given by \eqref{eq:gad}. All  PDFs have been estimated by generating $10^5$ samples.}
\label{ABCs}
\end{figure}

To estimate the different PDFs we have taken $K=10^3$ and  we have subsequently generated $\mathcal{N}=10^5$ $\Xi(t)$ matrices for the four different process. For each instance $s=1,\ldots,\mathcal{N}$ the area swept by each path is estimated as
\beeq{
A^{(s)}_{i}=\sum_{k=0}^K \lambda^{(s)}_i(k)\,,\quad\quad  i=1,\ldots, N\,.
}
where $\bm{\lambda}(k)=(-\lambda_{N}(k),\ldots,-\lambda_{1}(k),\lambda_{1}(k),\ldots,\lambda_{N}(k))$ are the eigenvalues of the matrix $\Xi$. This area is then rescaled to the $x$-variable $x^{(s)}_i=A^{(s)}_i/K^{3/2}$. Finally,  the sample set $\{x_i^{(s)}\}_{s=1}^{\mathcal{N}}$ for each area $i=1,\ldots,N$, is used to estimate the PDFs as:
\beeq{
\mathcal{P}_{N}^{(i)}(x)=\frac{1}{\mathcal{N}}\sum_{s=1}^{\mathcal{N}}\delta\left(x-x_i^{(s)}\right)\,\quad\quad i=1,\ldots,N\,.
}
Here we focus on the PDF  $Q_{N}(x)$ of the accumulated area.  An instance of the four processes can be found in figures \ref{fig2} and \ref{fig3}.\\
Results of the PDFs  estimated  by Monte Carlo simulations and comparison with the theoretical formulas for the PDF $Q_{N}(x)$ of the accumulated area are reported in figure \ref{ABCs} for both types of boundary conditions.

Looking at these figures some points are in order: the comparison between the  set of formulas \eqref{eq:gad} and \eqref{eq:gadR}  and Monte Carlo estimates could be done up to $N=10$ and $N=9$, respectively,  as otherwise the numerical evaluation of these exact formulas becomes prohibitively long. For the formulas \eqref{eq:gad2} and  \eqref{eq:gad2R} we have the an additional numerical problem due to the evaluation of the $B_N$ and, as a consequence, the comparison with Monte Carlo simulations is performed up to $N=2$ in both cases.

\section{Conclusions}
\label{sec:C}
In this work we have  generalised  the Airy distribution function of the  area swept by one Brownian particle performing a reunion to the case of the accumulated area swept by $N$ vicious Brownian particles. The exact formulas have been contrasted with Monte Carlo simulations showing perfect agreement.\\
There are several open problems which  we are currently investigating. First of all, whereas obtaining exact expressions for negative moments of the distribution of the accumulated area swept by $N$ vicious walkers is  rather simple, we wonder whether it is possible to obtain exacts expressions for the positive moments, perhaps generalising the techniques used for the case of one walker that can be found, for instance, in \cite{Takacs1991,Takacs1993,Kearney2007}. Secondly, we ponder whether it is possible to find exact formulas for the distribution of the area swept by either the bottom or the top paths. Within QMf this entails to being able to construct $N$-particle eigenfunctions for distinguishable fermions. As this seems to be a  daunting task, at the very least it would be interesting to study the properties of these distributions numerically. This  can be done fairly efficiently, in particular  for rate events, by combining the mapping to RMT we have used here together with the Wang-Landau algorithm,  a numerical technique that has been used in a similar context to study the statistics of extreme eigenvalues  \cite{Saito2010}. This analysis can obviously be extended to the cases in which the distribution of jumps of the Brownian motion  comes from either  thick or thin tails.

\begin{acknowledgements}
The authors warmly thank N. Kobayashi and M. Katori for email correspondence regarding the simulations. We also thank E. Barkai for pointing out some references.
\end{acknowledgements}

\appendix
\section{Taylor expansion of  Slater's determinant}
\label{ap:sd}
In this secion we discuss the Taylor expansion of a Slater determinant. Starting form the definition of the determinant and doing a Taylor expansion we write:
\beeq{
\det_{1\leq i,j\leq N}\varphi_{n_i}(z_j)&=\sum_{P\in S_{N}}\text{sign}(P)\prod_{i=1}^N\varphi_{n_P(i)}(z_i)\\
&=\sum_{P\in S_{N}}\text{sign}(P)\sum_{\ell_1,\ldots,\ell_N\geq 0}\frac{z_1^{\ell_1}\cdots z_N^{\ell_N}}{\ell_1!\cdots \ell_N!}\prod_{i=1}^N\varphi^{(\ell_i)}_{n_{P(i)}}(0)\\
&=\sum_{\ell_1,\ldots,\ell_N\geq 0}\frac{z_1^{\ell_1}\cdots z_N^{\ell_N}}{\ell_1!\cdots \ell_N!}\det_{1\leq i,j\leq N}\varphi^{(\ell_i)}_{n_{j}}(0)\,.
}
Due to the antisymmetric properties of the determinant, only those terms in the multiple sum with different values of $\ell$'s are different from zero. With this in mind we rewrite the sum over the $\bm{\ell}=(\ell_1,\ldots,\ell_N)$ as a sum over over $U_{N}$ of ordered indices (let's say $\ell_1<\ell_2<\cdots<\ell_N$) and then we sum over its permutation, \textit{viz.}
\beeq{
\det_{1\leq i,j\leq N}\varphi_{n_i}(z_j)&=\sum_{\ell_1,\ldots,\ell_N\geq 0}\frac{z_1^{\ell_1}\cdots z_N^{\ell_N}}{\ell_1!\cdots \ell_N!}\det_{1\leq i,j\leq N}\varphi^{(\ell_i)}_{n_{j}}(0)\\
&=\sum_{\bm{\ell}\in U_N}\frac{1}{\ell_1!\cdots \ell_N!}\sum_{P\in S_{N}} z_1^{\ell_{P(1)}}\cdots z_N^{\ell_{P(N)}}\det_{1\leq i,j\leq N}\varphi^{(\ell_{P(i)})}_{n_{j}}(0)\\
&=\sum_{\bm{\ell}\in U_N}\frac{1}{\ell_1!\cdots \ell_N!}\sum_{P\in S_N}\text{sign}(P) z_1^{\ell_{P(1)}}\cdots z_N^{\ell_{P(N)}}\det_{1\leq i,j\leq N}\varphi^{(\ell_{i})}_{n_{j}}(0)\\
&=\sum_{\bm{\ell}\in U_N}\frac{1}{\ell_1!\cdots \ell_N!}\det_{1\leq i,j\leq N}[z_i^{\ell_j}]\det_{1\leq i,j\leq N}[\varphi^{(\ell_{i})}_{n_{j}}(0)]\,,
}
as we wanted to show.

\section{Reunion}
\label{ap:nf}
For the normalisation factor we need to derive the propagator of $N$ vicious walkers in the semi-infinite line. This propagator is this case can be written as follows
\beeq{
G_{0}(\bm{y},T|\bm{x},0)=\frac{1}{T^{N/2}}\left(\frac{2}{\pi}\right)^N\int_0^\infty\cdots\int_0^\infty dq_1\cdots dq_N\Phi^{(0)}_{\bm{q}}\left(\frac{\bm{y}}{\sqrt{T}}\right)\Phi^{(0)}_{\bm{q}}\left(\frac{\bm{x}}{\sqrt{T}}\right) e^{-\frac{1}{2}\bm{q}^2}\,,
}
where $\Phi^{(0)}_{\bm{q}}(\bm{x})=(1/\sqrt{N!})\det_{1\leq i,j\leq N}[\sin(q_ix_j)]$. Using the Taylor expansion formula for the Slater determinant we arrive at
\beeq{
\Phi^{(0)}_{\bm{q}}(\bm{x})=\frac{1}{\sqrt{N!}}\sum_{\bm{s}\in U_{N}}\frac{1}{(2s_1+1)!\cdots(2s_N+1)!}\det_{1\leq i,j\leq N}[x_i^{2s_j+1}]\det_{1\leq i,j\leq N}[q_{i}^{2s_j+1}(-1)^{s_j}]\,,
}
with $\bm{s}=(s_1,\ldots,s_N)$. Noticing that the lowest contribution is obtained by of $s_i=i-1$, we obtain
\beeq{
\Phi^{(0)}_{\bm{q}}\left(\frac{\bm{x}}{\sqrt{T}}\right)&\simeq \beta_N T^{-\frac{N^2}{2}}\left[\prod_{i=1}^N x_i\right]\Delta_{N}\left(x_1^2,\ldots,x_N^2\right)\left[\prod_{i=1}^N q_i\right]\Delta_{N}(q_1^2,\ldots q_N^2)\,,
}
with $\beta_N=(-1)^{\frac{N(N-1)}{2}}/[ 1!\cdots(2N-1)!\sqrt{N!}]$. At this stage we notice that the propagator goes like
\beeq{
G_{0}(\bm{\epsilon},T|\bm{x},\epsilon)&\simeq \beta^2_N T^{-\frac{1}{2}N(2N+1)}\left[\prod_{i=1}^N \epsilon^2_i\right]\Delta^2_{N}\left(\epsilon_1^2,\ldots,\epsilon_N^2\right) B_N\,,\\
B_N&=\left(\frac{2}{\pi}\right)^{N}\int_0^{\infty}\cdots\int_0^\infty dq_1\cdots dq_Ne^{-\frac{1}{2}\bm{q}^2}\left[\prod_{i=1}^N q_i^2\right]\Delta_N^2(q_1^2,\cdots,q^2_N)
\label{eq:propapp}
}
where the constant $B_N$ is derived in appendix \ref{ap:constants}.\\
For reflecting boundary conditions the expression for the propagator is the same but with the Slater determinant given by $\Phi^{(0)}_{\bm{q}}(\bm{x})=(1/\sqrt{N!})\det_{1\leq i,j\leq N}[\cos(q_ix_j)]$. Using the expansion formula for the Slater determinant we arrive at
\beeq{
\Phi^{(0)}_{\bm{q}}(\bm{x}/\sqrt{T})=\frac{1}{\sqrt{N!}}\sum_{\bm{s}\in U_{N}}\frac{1}{(2s_1)!\cdots(2s_N)!}\det_{1\leq i,j\leq N}[T^{-s_{j}}x_i^{2s_j}]\det_{1\leq i,j\leq N}[q_{i}^{2s_j}(-1)^{s_j}]\,,
}
with $\bm{s}=(s_1,\ldots,s_N)$. Noticing that the lowest contribution is obtained by of $s_i=i-1$, we obtain
\beeq{
\Phi^{(0)}_{\bm{q}}\left(\frac{\bm{x}}{\sqrt{T}}\right)&\simeq \beta_N T^{-\frac{N(N-1)}{2}}\Delta_{N}\left(x_1^2,\ldots,x_N^2\right)\left[\prod_{i=1}^N q_i\right]\Delta_{N}(q_1^2,\ldots q_N^2)\,,
}
with $\beta_N=(-1)^{\frac{N(N-1)}{2}}/[ 0!\cdots (2(N-1))!\sqrt{N!}]$. At this stage we notice that the propagator goes like
\beeq{
G_{0}(\bm{\epsilon},T|\bm{x},\epsilon)&\simeq \beta^2_N T^{-\frac{N(2N-1)}{2}}\Delta^2_{N}\left(\epsilon_1^2,\ldots,\epsilon_N^2\right) F_N\,,\\
F_N&=\left(\frac{2}{\pi}\right)^{N}\int_0^{\infty}\cdots\int_0^\infty dq_1\cdots dq_Ne^{-\frac{1}{2}\bm{q}^2}\Delta_N^2(q_1^2,\cdots,q^2_N)
}
where the constant $F_N$ is derived in appendix \ref{ap:constants}.

\section{Meander}
\label{app:meander}
For the case of a meander of  $N$ vicious walkers (a star with a wall) we need to integrate over the final position $\bm{x}$.  Using the Taylor expansion of the Slater determinant as explained in the main text we can write
\beeq{
\int_{\bm{W}_N} d^N\bm{x} G_N^{(1)}(\bm{x},T|\bm{\epsilon},0)&=\sum_{\bm{n}}\int_{\bm{W}_N} d^N\bm{x}\,\Phi^{(1)}_{\bm{n}}(\bm{x})\overline{\Phi}^{(1)}_{\bm{n}}(\bm{\epsilon})e^{-E^{(1)}_{\bm{n}} T}\\
&\simeq (2\lambda)^{\frac{N^2}{3}}\gamma_N\left[\prod_{i=1}^N \epsilon_i\right]\Delta_N(\epsilon_1^2,\ldots \epsilon_N^2)\,\sum_{\bm{n}}B_N(\alpha_{n_1},\ldots,\alpha_{n_N})\Delta_N(\alpha_{n_1},\dots \alpha_{n_N})e^{-E^{(1)}_{\bm{n}} T}\,,
}
where we have defined
\beeq{
B_N(\alpha_{n_1},\ldots,\alpha_{n_N})&=\frac{1}{\prod_{i=1}^N\text{Ai}'(-\alpha_{n_i})}\int_{\bm{W}_N} d^N\bm{x}\det_{1\leq i,j\leq N}\left[\text{Ai}(x_j-\alpha_{n_i})\right]\,.
}
Similarly, the propagator in the  denominator goes like
\beeq{
\int_{\bm{W}_N} d^N\bm{x} G_N^{(0)}(\bm{x},T|\bm{\epsilon},0)&\simeq T^{-\frac{N^2}{2}}\gamma_N F_{N}\left[\prod_{i=1}^N \epsilon_i\right]\Delta_N(\epsilon_1^2,\ldots \epsilon_N^2)
}
where $F_N$ is a constant with no simple expression.\\
For reflecting boundary conditions we have instead the following expression for the numerator
\beeq{
&\int_{\bm{W}_N} d^N\bm{x}G_{N}^{(1)}(\bm{x},T|\bm{\epsilon},0)=\sum_{\bm{n}}\int_{\bm{W}_N} d^N\bm{x}\Psi_{\bm{n}}^{(1)}(\bm{x})\overline{\Psi}_{\bm{n}}^{(1)}(\bm{\epsilon})e^{-E_{\bm{n}}^{(1)}T}\\
&=\frac{\delta_N}{N!}\Delta_{N}(x_1^2,\ldots,x_N^2)\sum_{\bm{n}}\frac{\Delta_N(\beta_{n_1},\ldots,\beta_{n_N})}{\beta_{n_1}\cdots\beta_{n_N}}e^{-E_{\bm{n}}^{(1)}T}(2\lambda)^{\frac{N(2N-1)}{6}}(2\lambda)^{N/6}\\
&\int_{\bm{W}_N} d^N\bm{x}\frac{1}{\prod_{i=1}^N\text{Ai}(-\beta_{n_i})}\det_{1\leq i,j\leq N}[\text{Ai}((2\lambda)^{1/3}x_j-\beta_{n_i})]\\
&=\frac{\delta_N}{N!}\Delta_{N}(x_1^2,\ldots,x_N^2)(2\lambda)^{\frac{N(2N-1)}{6}-N/6}\sum_{\bm{n}}\frac{\Delta_N(\beta_{n_1},\ldots,\beta_{n_N})}{\beta_{n_1}\cdots\beta_{n_N}}e^{-E_{\bm{n}}^{(1)}T}\\
&\int_{\bm{W}_N} d^N\bm{z}\frac{1}{\prod_{i=1}^N\text{Ai}(-\beta_{n_i})}\det_{1\leq i,j\leq N}[\text{Ai}(z_j-\beta_{n_i})]\\
&=\frac{\delta_N}{N!}\Delta_{N}(x_1^2,\ldots,x_N^2)(2\lambda)^{\frac{N(N-1)}{3}}\sum_{\bm{n}}\frac{\Delta_N(\beta_{n_1},\ldots,\beta_{n_N})}{\beta_{n_1}\cdots\beta_{n_N}}e^{-E_{\bm{n}}^{(1)}T}\\
&\frac{1}{\prod_{i=1}^N\text{Ai}(-\beta_{n_i})}\int_{\bm{W}_N} d^N\bm{z}\det_{1\leq i,j\leq N}[\text{Ai}(z_j-\beta_{n_i})]
}
\section{On the normalisation constants}
\label{ap:constants}
Here we derive the expressions for the two normalisation constants for the case of reunions for absorbing and reflecting boundary conditions. For both cases we recall the well-known result of Selberg's integral \cite{forrester2010log}:
\beeq{
\frac{1}{N!}\int_0^\infty\cdots\int_0^{\infty}|\Delta(x_1,\ldots,x_N)|^{2c}\prod_{i=1}^Nx_i^{a-1}e^{-x_i} dx_i=\prod_{j=0}^{N-1}\frac{\Gamma(a+jc)\Gamma((j+1)c)}{\Gamma(c)}
}
For the constant $B_N$  we do the change of variables $q_i^2/2=y_i$ so that $q_i dq_i=dy_i$ or $dq_i=\frac{dy_i}{\sqrt{2y_i}}$, so we can write
\beeq{
B_N&=\left(\frac{2}{\pi}\right)^{N}\int_0^{\infty}\cdots\int_0^\infty dq_1\cdots dq_Ne^{-\frac{1}{2}\bm{q}^2}\left[\prod_{i=1}^N q_i^2\right]\Delta_N^2(q_1^2,\cdots,q^2_N)\\
&=\left(\frac{2}{\pi}\right)^{N}2^{-N/2} 2^{N} 2^{N(N-1)}\int_0^{\infty}\cdots\int_0^\infty dy_1\cdots dy_N\prod_{i=1}^Ne^{-y_i} y_i^{-1/2}\left[\prod_{i=1}^N y_i\right]\Delta_N^2(y_1,\cdots,y_N)\\
&=\frac{2^{\frac{1}{2}N(2N+1)}}{\pi^N}N!\prod_{j=0}^{N-1}\Gamma\left(\frac{3}{2}+j\right)\Gamma(j+1)=\frac{2^{\frac{1}{2}N(2N+1)}}{\pi^N}\prod_{j=0}^{N-1}\Gamma(2+j)\Gamma\left(\frac{3}{2}+j\right)
\label{eq:39}
}
Similarly for the case of reflecting boundary conditions we have
\beeq{
F_N&=\left(\frac{2}{\pi}\right)^{N}\frac{1}{2^{N/2}}2^{N(N-1)}\int_0^{\infty}\cdots\int_0^\infty dy_1\cdots dy_N\Delta_N^2(y_1,\cdots,y_N)\prod_{i=1}^Ny_i^{-1/2} e^{-y_i}\\
&=N!\frac{2^{\frac{N(2N-1)}{2}}}{\pi^N}\prod_{j=0}^{N-1}\Gamma\left(\frac{1}{2}+j\right)\Gamma\left(j+1\right)=\frac{2^{\frac{N(2N-1)}{2}}}{\pi^N}\prod_{j=0}^{N-1}\Gamma\left(\frac{1}{2}+j\right)\Gamma\left(j+2\right)
}

\section{Derivatives of $F(x,\gamma)$}
\label{app:dF}
Starting from the following expression  for $F(x,\gamma)$:
\beeq{
F(x,\gamma)=\frac{\sqrt{3}}{x\sqrt{\pi}}u^{2/3}(x,\gamma)e^{-u(x,\gamma)}U\left(\frac{1}{6},\frac{4}{3},u(x,\gamma)\right)\,,\quad u(x,\gamma)=\frac{2\gamma^3}{27x^2}\,,
}
and using the following property
\beeq{
x\frac{\partial U(a,b,x)}{\partial x}=(a-b+x)U(a,b,x)-U(a-1,b,x)\,,
\label{eq:proper}
}
we notice that we can write the expression
\beeq{
\frac{\partial^n F(x,\gamma)}{\partial x^n}=\frac{\sqrt{3}}{x^{n+1}\sqrt{\pi}}u^{2/3}(x,\gamma)e^{-u(x,\gamma)}\sum_{\ell=0}^n C^{(n)}_{\ell}U\left(\frac{1}{6}-\ell,\frac{4}{3},u(x,\gamma)\right)\,.
\label{eq:res1}
}
for some set of coefficients  $\{C^{(n)}_\ell\}$ still to be determined. The expression \eqref{eq:res1} is certainly correct for $n=1$ and $n=2$. Let us them assume is holds for any $n$ and performe one more derivative with respect to $n$:
\beeq{
\frac{\partial^{n+1} F(x,\gamma)}{\partial x^{n+1}} &=\frac{1}{x^{n+2}\sqrt{3\pi}}u^{2/3}(x,\gamma)e^{-u(x,\gamma)}\\
&\sum_{\ell=0}^n C^{(n)}_{\ell}\left[(-7-3n+6u(x,\gamma))U\left(\frac{1}{6}-\ell,\frac{4}{3},u(x,\gamma)\right)-6u(x,\gamma)U'\left(\frac{1}{6}-\ell,\frac{4}{3},u(x,\gamma)\right)\right]\,.
}
But using  the  property \eqref{eq:proper}, we can write
\beeq{
u(x,\gamma)U'\left(\frac{1}{6}-\ell,\frac{4}{3},u(x,\gamma)\right)=\left(\frac{1}{6}-\ell-\frac{4}{3}+u\right)U\left(\frac{1}{6}-\ell,\frac{4}{3},u(x,\gamma)\right)-U\left(\frac{1}{6}-\ell-1,\frac{4}{3},u(x,\gamma)\right)\,.
}
Gathering results we find
\beeq{
\frac{\partial^{n+1}F(x,\gamma)}{\partial x^{n+1}}&=\frac{\sqrt{3}}{x^{n+2}\sqrt{\pi}}u^{2/3}(x,\gamma)e^{-u(x,\gamma)}\\
&\sum_{\ell=0}^n C^{(n)}_{\ell}\left[(-n +2\ell)U\left(\frac{1}{6}-\ell,\frac{4}{3},u(x,\gamma)\right)+2U\left(\frac{1}{6}-\ell-1,\frac{4}{3},u(x,\gamma)\right)\right]\,.
}
On the  other hand, we want to write this result as \eqref{eq:res1} for $n\to n+1$. This implies to rewrite the sum as:
\beeq{
&\sum_{\ell=0}^n C^{(n)}_{\ell}\left[(-n +2\ell)U\left(\frac{1}{6}-\ell,\frac{4}{3},u(x,\gamma)\right)+2U\left(\frac{1}{6}-\ell-1,\frac{4}{3},u(x,\gamma)\right)\right]\\
&=\sum_{\ell=0}^n C^{(n)}_{\ell}(-n +2\ell)U\left(\frac{1}{6}-\ell,\frac{4}{3},u(x,\gamma)\right)+2\sum_{\ell=0}^k C^{(n)}_{\ell}U\left(\frac{1}{6}-\ell-1,\frac{4}{3},u(x,\gamma)\right)\\
&=\sum_{\ell=0}^n C^{(n)}_{\ell}(-n +2\ell)U\left(\frac{1}{6}-\ell,\frac{4}{3},u(x,\gamma)\right)+2\sum_{\ell=1}^{n+1} C^{(n)}_{\ell-1}U\left(\frac{1}{6}-\ell,\frac{4}{3},u(x,\gamma)\right)\\
&=\sum_{\ell=0}^{n+1} C^{(n+1)}_{\ell}U\left(\frac{1}{6}-\ell,\frac{4}{3},u(x,\gamma)\right)\,.
}
This results in the following set of recurrence relations for the set of coefficents $\{C_{\ell}^{(n)}\}$:
\beeq{
C^{(n+1)}_0&=-nC_{0}^{(n)}\,,\\
C_{\ell}^{(n+1)}&=C_{\ell}^{(n)}(-n+2\ell)+2C_{\ell-1}^{(n)}\,,\quad \ell=1,\ldots,n\,,\\
C_{n+1}^{(n+1)}&=2C_{n}^{(n)}\,,
}
with the initial condition $C_0^{(0)}=1$. By checking explicitly the value of some of these coeffcients, and with the help of the Sloane database\footnote{At \texttt{https://oeis.org}}, we arrive at the solution:
\beeq{
C^{(n)}_\ell=\frac{n!}{2^{n-2\ell} (n-\ell)!(2\ell-n)!}\,,\quad \ell=0,\ldots,n\,.
}
A Similar analysis can be perform ny doing derivatives with respect to $\gamma$. Starting from \eqref{eq:res1} one notices that
\beeq{
\frac{\partial^{k+n} F(x,\gamma)}{\partial \gamma^k\partial x^n}=\frac{\sqrt{3}}{x^{n+1}\sqrt{\pi}\gamma^{k}}u^{2/3}(x,\gamma)e^{-u(x,\gamma)}\sum_{\ell=0}^{n}\sum_{s=0}^{k}C^{(n)}_{\ell} D^{(k)}_{s}(\ell)U\left(\frac{1}{6}-\ell-s,\frac{4}{3},u(x,\gamma)\right)\,,
\label{eq:res2b}
}
for some set of coefficients  $\{D_s^{(k)}(\ell)\}$. Performing one more derivative with respect to $\gamma$ allows us to arrive at the following set of recurrence relations for those, \textit{viz.}
\beeq{
D_0^{(k+1)}(\ell)&=-\left(\frac{3}{2}+k+3\ell\right)D_{0}^{(k)}(\ell)\,,\\
D_{s}^{(k+1)}(\ell)&=-\left(\frac{3}{2}+k+3\ell+3s\right)D_{s}^{(k)}(\ell)-3D_{s-1}^{(k)}(\ell)\,,\quad\quad s=1,\ldots,k\,,\\
D_{k+1}^{(k+1)}(\ell)&=-3D_{k}^{(k)}(\ell)\,,
}
with the initial condition $D_0^{(0)}(\ell)=1$. As we  are only interested in the orders $k=0$ and $k=2$, we do not need a general solution. This yields
\beeq{
D_0^{(0)}(\ell)&=1\,,\\
D_{0}^{(1)}(\ell)&=-\frac{3}{2}\left(1+2\ell\right)\,,\quad\quad D_{1}^{(1)}(\ell)=-3\,\\
D_0^{(2)}(\ell)&=\frac{3}{4} (1 + 2\ell) (5 + 6\ell)\,,\quad\quad D_1^{(2)}(\ell)=3(7+6\ell)\,,\quad\quad D_2^{(2)}(\ell)=9\,.
}

\section{Moments}
\label{app:moments}
We first recall the following identity
\beeq{
\int_0^{\infty} ds s^{\mu} e^{-xs}=\frac{\Gamma(\mu+1)}{x^{1+\mu}}\,.
}
From here we write
\beeq{
M_{-(1+\mu)}&=\int_0^\infty dx x^{-(1+\mu)} F(x)\\
&=\frac{1}{\Gamma(\mu+1)}\int_0^\infty dx F(x)\int_0^{\infty} ds s^{\mu} e^{-xs}\\
&=\frac{1}{\Gamma(\mu+1)}\int_0^{\infty} ds s^{\mu}\int_0^\infty dx F(x) e^{-xs}\\
&=\frac{1}{\Gamma(\mu+1)}\int_0^{\infty} ds s^{\mu}\widehat{F}(s)\\
&=\frac{1}{\Gamma(\mu+1)}\int_0^{\infty} ds s^{\mu+E}e^{-as^{2/3}}
}
But
\beeq{
\int_0^{\infty} ds s^{\mu+E}e^{-as^{2/3}}&=\int_0^{\infty} ds (s^{2/3})^{\frac{3}{2}(\mu+E)}e^{-as^{2/3}}\\
&=\frac{3}{2}\int_0^{\infty} dy y^{\frac{3}{2}(\mu+E)+1/2}e^{-a y}\\
&=\frac{3}{2}a^{-\frac{3}{2}(\mu+E+1)}\int_0^{\infty} dy y^{\frac{3}{2}(\mu+E)+1/2}e^{- y}\\
&=\frac{3}{2}a^{-\frac{3}{2}(\mu+E+1)}\Gamma\left(\frac{3}{2}(\mu+E+1)\right)
}
Thus denoting $\nu=1+\mu$ we finally have
\beeq{
M_{-\nu}=\frac{3}{2}a^{-\frac{3}{2}(\nu+E)}\frac{\Gamma\left(\frac{3}{2}(\nu+E)\right)}{\Gamma(\nu)}
}

\bibliographystyle{spmpsci}      % mathematics and physical sciences

\bibliography{bib}   % name your BibTeX data base

\end{document}